\documentclass[intlimits,twoside,a4paper]{article}
\usepackage[cp1251]{inputenc}
\usepackage[eqsecnum]{cmpj3}
\usepackage{graphicx}

\usepackage{color}
\usepackage{subfigure}
\usepackage{bm}
\usepackage{soul}
\usepackage{float}
\usepackage{hyperref} 

\usepackage{enumerate}

\issue{2021}{24}{3}{33401}
\doinumber{10.5488/CMP.24.33401}

\title [Molecular dynamics of Janus dumbbells and spherical particles]  %
{ 
Molecular Dynamic study of model two-dimensional systems involving Janus 
dumbbells and spherical particles}

\author{\L. Baran, K. D\k{a}browska,
W. R\.{z}ysko, S. Soko\l owski}

\address{Department for Theoretical Chemistry, Faculty of Chemistry,
Maria Curie-Sk{\l}odowska University,  Lublin 20-031, Poland}

\date{Received June 17, 2020, in final form May 25, 2021}

\begin{document}
	
\maketitle

\begin{abstract}
	
We have performed an extensive constant temperature Molecular Dynamics study of
two-dimensional systems involving Janus dumbbells and spherical particles.
Janus dumbbells have been modelled as two spheres, labeled 1 and 2, joined together
via harmonic bonds. Sphere 1 of a selected Janus dumbbell attracts 
the spheres of the same kind 
on other Janus
dumbbells, while the interactions between the pairs 1-1 and 1-2 were repulsive.
On the other hand, the spherical particles are attracted by centers 2 and repelled by
the centers 1 of Janus particles.
We have shown that the structure of
oriented phases that can be formed in the system depends on the bond length of Janus 
dumbbells and the ratio of the number of spherical particles to the number of 
Janus dumbbells in the system. 
The presence of spherical particles
is necessary to develop oriented phases. For the assumed model, the formation of 
oriented phases in the system depends on the concentration of spherical particles. 
Equal numbers of Janus and spherical particles create optimal conditions for the 
formation of lamellar phases. 

\end{abstract}

\section{Introduction}

Particles of anisotropic shape and interactions
play a significant role in the development of 
different self-assembling structures~\cite{i1,i2,i3}.
The self-assembly of these particles, which are often
called as the ``building blocks'', has now emerged 
as a versatile platform  for the creation of 
novel materials with requested properties.
These new materials have found  numerous applications in science,
 medicine, biotechnology, for the production of photonic crystals, 
 detection of biomolecules, energy harvesting
 and in many other fields~\cite{i4,i5,i6,i7,i8,i9,i10,i11,i12,i13}.
On the other hand, the investigation of self-organization processes 
has also led to significant progress in the preparation and design
of anisotropic particles of precisely defined size, shape, and functionality.

Janus particles that are
composed of  two or, in some cases, of a greater number of  chemically distinct parts, 
constitute a unique class of anisotropic particles. 
Among different types of Janus-like particles, the so-called Janus dumbbells
that are built of two jointed spheres of different chemical character are 
one of the simplest ``building blocks'' and have received much attention 
recently. In particular, the behavior and self-assembly
of Janus dumbbells in bulk two- and three-dimensional phases,
as well as at different interfaces have been extensively
studied using computer simulations and theoretical approaches, e.g., 
density functional 
methods~\cite{i15,i16,i17,i18,i19,i19a,i20,i21,i22}. 
Theoretical methods  for the description of  the systems of anisotropic nanoparticles are  
similar to the methods used in the studies of systems  
involving amphiphilic molecules, e.g., surfactants, or even to some extent,
of dipolar molecules~\cite{j1,j2}.

The conditions of self-organization and morphology of emerging phases
formed by anisotropic nanoparticles  depend on the presence in the system
of other, usually spherical species~\cite{j3,j4,j5,j6,j66,j7}.
Similar effects to those observed for nanoparticles
have also occurred on molecular length scales, namely in the case of  
formation of supramolecular structures 
in mixtures involving anisotropic and spherical molecules~\cite{j8,j9}.
In both cases, i.e., for nanoparticles, and for molecules organizing into
supramolecules, the
properties of self-assembled structures  depend on     
the system composition and on the ratio of sizes of the involved components.
Careful tuning of these
parameters can enable directing the system self-assembly toward
architectures with predefined morphology and functions.

Self-assembling nanoparticles are  dissolved in an inert solvent 
that does not directly participate in the resulting structures.
The main problem with accurate modelling of those systems 
at a microscopic level is connected with the difference in sizes between 
nanoparticles and solvent molecules.
In experimental setups, the size of nanoparticles 
is usually bigger than 100~nm, i.e., the size of nanoparticles is at least 
a few hundred times greater than the size of the fluid
molecules.  
Such a large difference can be handled
using thermodynamic approaches that are
based on macroscopic quantities such as surface and line
tensions, but then the microscopic structure of 
the system is neglected~\cite{s3,s4}.  
In the case of computer simulation
that is focused on a detailed description
of the microscopic structure, 
the difference in sizes between fluid and nanoparticles should be 
much smaller. In fact, in the existing publications, 
the size of simulated nanoparticles  
has  ranged from a few to a dozen or so diameters 
of the fluid solvent molecules~\cite{s6,s7,s8,s9}.
Thus, the simulated nanoparticles were at least ten times smaller 
than in experimental systems.

One of the possible methods of simulating systems
containing nanoparticles is the use of the so-called implicit solvent
models~\cite{s1,s2}.
According to these models, 
the presence of solvent molecules is neglected.
The interactions between all remaining species are
treated as being ``effective'' or solvent-mediated. 
The term ``effectiveness'' means that the interaction potentials take into account
the presence of solvent molecules in an implicit way.
The implicit solvent models are  computationally 
efficient and 
can provide a reasonable description of 
the self-organization effects.

The main focus of the present study is on the 
presentation of the simulation results
of the development
of self-organized phases in two-dimensional systems involving Janus dumbbells
of different elongation and  different concentrations of 
spherical particles.
We want to study how the two factors, namely, elongation of dimers and concentration
of spherical particles,
affect the formation of self-assembled structures.
The systems are modelled within the 
implicit solvent model.
The structural properties  of investigated mixtures have also been 
contrasted with the properties of systems involving Janus dumbbells alone. 
The systems of homo- and heteronuclear dimers 
of different elongation have already
been investigated in numerous works~\cite{dod1,dod1a,dod2,dod2a,dod3,dod4},
but the models of interactions used in the above-cited publications
differ from those used by us.

We employ Molecular Dynamics simulations to
investigate the occurrence of ordered phases 
at different
concentrations of spherical particles and at different temperatures.
However, the aim of the work is not to determine the complete phase diagrams, 
but rather to draw attention to the possibility 
of the formation of different structures under selected conditions.
The studies of similar  mixtures have already been
carried out~\cite{rev1,rev2,rev3}, but they  concentrated
on the aggregation of colloidal spheres mediated by Janus dimers.

The paper is organized as follows. In the following
section (section~\ref{sec:model}) we present the model. Then, the simulation method
(section~\ref{sec:sim}) is outlined.
In section~\ref{sec:rd} we discuss the obtained results, starting from the systems
in which the Janus dumbbells are built of two tangentially jointed
spheres and different amounts of spherical particles. 
Next, we proceed to systems with shorter Janus dimers.
Finally, we consider systems involving only Janus particles,
without the spherical ones. The latter systems resemble
those studied by Bordin and Krott~\cite{i19,i19a}. The main results
are concluded in section~\ref{sec:con}.

This work is dedicated to  
Yuri Kalyuzhnyi, a distinguished scientist in the field of statistical thermodynamical theory of liquids, on
behalf of his 70th birthday.  We highly appreciate his friendship during the last decades
and we would like to thank him for many interesting and fruitful scientific discussions.

\section{Model}\label{sec:model}

We consider systems involving either pure 
Janus dumbbells or mixtures of Janus dumbbells and spherical 
(circular) particles. 
Each Janus dumbbell is built of
two jointed spheres of the same diameter, $\sigma$, located at a distance $d$ apart. 
The atoms constituting Janus
dumbbells are distinguished by the indices 1 and 2, whereas the
index  3 refers to spherical particles.

The binding between atoms 1 and 2 is ensured by the harmonic potential

\begin{equation}\label{eq:bond}
 U_b(r)=k_H(r-d)^2,
\end{equation}
where $k_H$ is a constant.
The interactions 
between all pairs $ij$, ($i,j=1,2,3$), are described 
by the potential

\begin{equation}\label{eq:LJ}
 u_{ij}(r)=\left\{
 \begin{array}{ll}
  \Phi_{ij}^{(I)}(r)- \Phi_{ij}^{(I)}(r_{ij,c}) - (r-r_{ij,c})\left. \frac{\rd \Phi_{ij}^{(I)}(r)}{\rd r}\right| & r\leqslant r_{ij,c}  \\
  0, & r>r_{ij,c},
 \end{array}
\right.
\end{equation}
where 
\begin{equation}\label{eq:LJ1}
 \Phi_{ij}^{(I)}(r)=4\varepsilon_{ij}[(\sigma_{ij}/r)^{12}-(\sigma_{ij}/r)^{6}].
\end{equation}
This potential combines the standard Lennard-Jones(12,~6)
function and subtracts a linear term based on the cutoff distance, $r_{ij,c}$, so that both,
the potential and the force, tend continuously to zero at the cutoff~\cite{Toxvaerd}.
In the above, $\varepsilon_{ij}$ and $\sigma_{ij}=0.5(\sigma_i+\sigma_j)$ are the energy and the size parameters
for the interactions between the pairs of the type $i$ and $j$.
The Lennard-Jones sizes of the atomic species 1 and 2 were equal, $\sigma_1=\sigma_2\equiv \sigma$, 
and $\sigma$ was used as the length unit. The unit of the energy, however, was $\varepsilon_{22}\equiv\varepsilon$.
The size of the particles 3 was two times smaller than $\sigma$, $\sigma_3=\sigma/2$. The diameter 
$\sigma_3$ was ad hoc selected, though it is similar to that used in~\cite{rev1,rev2}.

The reduced
quantities are distinguished by asterisks, i.e., $d^*=d/\sigma$, 
$\sigma^*_i=\sigma_i/\sigma$ and $\varepsilon_{ij}^*=\varepsilon_{ij}/\varepsilon$,
thus $\sigma_3^*=0.5$.
The reduced temperature is defined as usual,
$T^*=k_{\rm{B}}T/\varepsilon$ where $k_{\rm{B}}$ is the Boltzmann constant.

If the cut-off distance equals $r_{ij,c}=2^{1/6}\sigma_{ij}$, then the interaction potential
is entirely repulsive~\cite{wca,hess}.
According to the assumed model, the attractive interactions exist only between the pairs $11$ and $23$.
In other words, the attractive forces exist only between attractive parts of Janus particles
and the repulsive parts of Janus particles and spherical molecules.
All the remaining interactions are repulsive. The energy parameters, together with the cut-off distances 
are collected in table~\ref{tb1}. 

\begin{table}[htb]
\caption{The values of the parameters of equations~(\ref{eq:LJ}) and  (\ref{eq:LJ1}).}
\label{tb1}
\begin{center}
	\vspace{3pt} \noindent
\begin{tabular}{|l|l|l|l|l|l|l|l|l|l|l|l|}
 \hline
  $\varepsilon_{11}^*$ & $\varepsilon_{22}^*$ & $\varepsilon_{33}^*$& $\varepsilon_{12}^*$&
 $\varepsilon_{13}^*$ & $\varepsilon_{23}^*$ & $r_{11,c}^*$ &  $r_{23,c}^*$ & $r_{22,c}^*$ &$r_{12,c}^*$
 &$r_{13,c}^* $ & $r_{33,c}^* $
 \\
 \hline
  4 & 1 & 1 & 1 & 1 & 4 & 1.6 & $1.6\sigma_{23}^*$ & $2^{1/6}$& $2^{1/6}$ & $2^{1/6}\sigma_{13}^*$ &
 $2^{1/6}\sigma_{33}^*$ 
  \\
 \hline
\end{tabular}
	\vspace{2pt}
\end{center}
\end{table}

The depth of the attractive well of the Lennard-Jones energy $u_{11}(r)$ 
is the same as for the pair of particles  2 and 3. Instantaneously, the particles 3
are smaller than the centers 1 and 2. Therefore, under certain conditions we can expect
the formation of  structures with the ``building blocks''  involving particles~3 (cf. reference~\cite{dod2a}).
To speed up the simulations, the range of all attractive interactions 
has been assumed to be short. 

The model investigated by us does not describe any specific experimental setup.
Similarly to the works~\cite{dod2,dod2a,dod3,dod4} we aimed at finding sets of 
parameters for which the development of interesting structures
 could be observed. 
Before selecting the values of the parameters from table~\ref{tb1},
we carried out several auxiliary short simulation runs.

We stress that the harmonic bonding potential constant [cf. equation~(\ref{eq:bond})] 
was $k_H/\sigma^2=1000\varepsilon$. A high value of this constant ensures very small
harmonic vibrations of the length of Janus dumbbells bonds. 

\section{Simulation details}\label{sec:sim}

The simulations were carried out using the LAMMPS
package~\cite{LAMMPS}. The in-plane, 0$XY$, cell dimension was $L\times L$
and the periodic boundary conditions were applied. The numbers of Janus dumbbells and
spherical particles are designated as $N_J$ and $N_3$, respectively.
The system initial composition is given in terms of the ratio
 $\chi=N_3/N_J$.
The numbers $N_J$ and $N_3$, as well the ratio
$\chi$ always refer to the nominal values, averaged over the
entire system. 

To model the particles immersed in an implicit solvent,
a Langevin thermostat was employed. 
The masses of all atomic components were set to 1. The Langevin thermostat is weakly coupled to the system, 
the dumping constant in the LAMMPS
input file~\cite{LAMMPS_Lang} was 10. Since the aim of the simulations was
to study the systems at equilibrium, the selection of masses of the particles was
irrelevant.

The reduced
time step, $\tau=t\sqrt{\varepsilon/m\sigma^2}$ (where $t$ is the time step in seconds),
was $\tau^*=0.001$. The choice of $\tau^*$ is a compromise between the time 
and accuracy of simulations. Our selection
of $\tau^*$ resulted from the performed tests. Unfortunately, during simulations, we determined only the values of $S$.

Before performing the final runs, we considered different possibilities for selecting the order parameter that can be easily computed in simulations and decided to select the nematic order parameter of $S$.
We modified the description of equation~(\ref{eq:3_1}) and the description of the results. We also added several references related to this problem. 

A great majority of the results are for $N_J=900$, but to verify 
the obtained results, several calculations were also performed
for $N_J=1800$. The total number of 
simulated spherical entities was $2N_J+N_3 = 2N_J+\chi N_J$.

In the forthcoming discussion, the nominal system  density is 
given by the total packing fraction. It is defined as
$\rho^*=(N_JV_J+N_3V_3)/V_{\rm{box}}$, where $V_J=\piup\sigma^3/3-
\piup(2\sigma+d)(\sigma-d)^2/12$
is the volume of the Janus dumbbell, 
$V_3$ is the volume of a spherical particle and $V_{\rm{box}}=L^2\sigma$.

We carried out equilibration runs for $5\times 10^7$ time steps and after that 
time the production steps were continued until  sufficient 
statistics for the measured properties were reached.
At the initial stage of equilibration,
the nominal  density of every system $\rho^*$ was adjusted by short NPT runs
during which the size of the simulation cell was varied. After that, 
the size of the cell and thus the nominal system density,  $\rho^*$,  were kept constant
during consecutive NVT simulation.

The purpose of the simulation was to coarsely determine the 
phase diagrams using the method of block density analysis.
Our calculations were carried out for
the nominal values of $\chi$  equal to $\chi=1, 2$ and 4.

The values of  $\rho^*$
were selected in such a way that they were approximately equal to
the average density of the coexisting phases at the lowest temperature.
The selection of specific values of $\rho^*$
was made by conducting preliminary NVT simulations for various nominal
densities. 

The formation of ordered phases can result in different
numbers of particles of different kind and different 
compositions in some parts of the systems. 
In the forthcoming discussion, the symbols $\rho^*_L$ and $\chi_L$
refer to distinguished parts (phases) of the system.
 
To characterize the systems, the following structural 
characteristics were evaluated:
\renewcommand{\theenumi}{\roman{enumi}}
\begin{enumerate}
\item radial distribution functions between the species 1, 2, 3 and the radial distribution function between centers of mass of Janus dumbbells, particles,
\item the distribution of cluster sizes formed by Janus dumbbells,
\item the order parameter for characterizing the orientational ordering of Janus dumbbells.
\end{enumerate}

From the center of mass radial distribution function, the distance that corresponds to its
first minimum, $R_{JJ,m}$, 
was estimated.
 This distance was  in turn used to calculate the cluster
size distributions using the algorithm of Rapaport~\cite{Rapaport}.

The development of ordered structues in
the systems under study resembles the behavior of liquid crystals.
The order in liquid crystals can be studied by introducing 
different order parameters~\cite{AAA1,AAA2}, as well as the complex positional order parameters~\cite{AAA3}.
Since the analysis of preliminary data indicated the appearance
of lamellar phases under certain conditions, we decided to characterize the
changes of the orientational ordering of Janus dumbbells by
the tensor order parameter. Its definition is
given in the work of Eppenga and Frenkel~\cite{AAA}.
It reads 
\begin{equation}
Q_{\alpha\beta}=\frac{1}{N_J} \sum_i^N \left [ 2b_{\alpha}(i)b_{\beta}(i)-\delta_{\alpha \beta} \right ].
\label{eq:3_1}
\end{equation} 
In the above, $b_{\alpha}(i)$ [$b_{\beta}(i)$] is the $\alpha$-th ($\beta$-th) coordinate of the unit vector $\mathbf{b}$,
 specifying the orientation of the molecule $i$, and $\delta_{\alpha \beta}$ 
 is the Kronecker delta function. {\bf Q}
 is a traceless symmetric second rank tensor, with three eigenvalues $\lambda_+$, $\lambda_0$,
 $\lambda_-$.
 The nematic order parameter $S$  is defined as the largest positive eigenvalue, $\lambda_+$. 
 $S$ takes on the values between 0 and 1 and equals 0 in completely
 disordered phases and equals 1 in perfectly ordered phases, respectively~\cite{AAA}.

 \section{Results and  discussion}\label{sec:rd}

The principal aim of our study was to investigate the changes in the structure of the system with the change
of the length of Janus dumbbells and with the composition of the system. 
We begin our discussion with the presentation of the results for the systems involving Janus dumbbells  (of the bond length $d^*=1$) and spherical particles at
three nominal concentrations, $\chi=1$, 2 and 4. The calculations were carried out
at several temperatures from the range $0.2\leqslant T^*<0.5$. 
The total nominal system density was adjusted to $\rho^*=0.217$.

\begin{figure}[!ht]
  \centering
\includegraphics[scale=0.24]{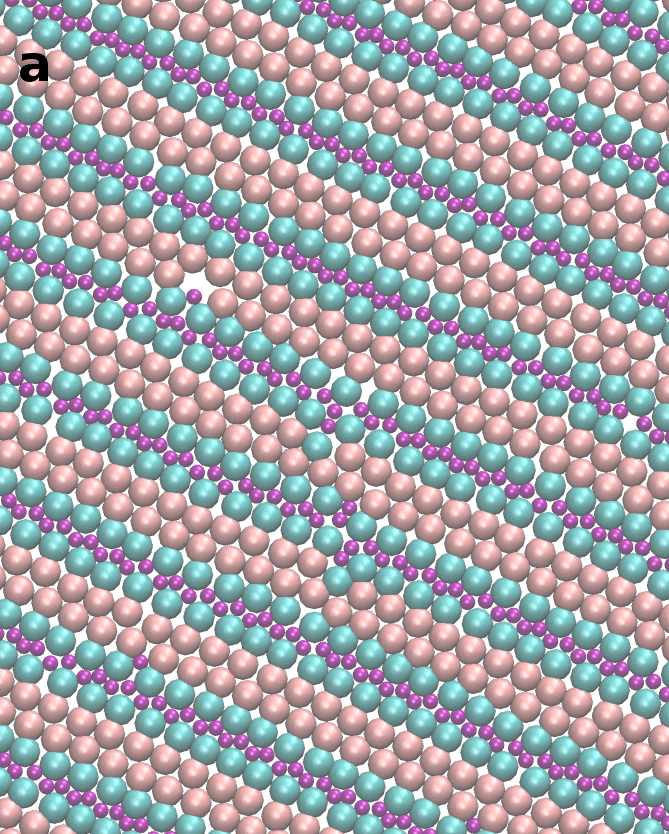}
\includegraphics[scale=0.63]{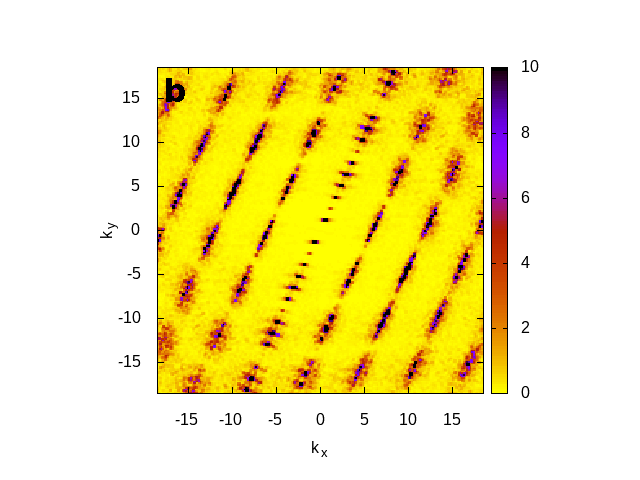}
\caption{ 
(Colour online) Panel a shows the snapshot of a panel of the lamellar phase and
panel b --- the two-dimensional structure factor obtained from
 $g_{11}(r)$. The calculations are at $T^*=0.2$. The nominal
 density  is $\rho^*=0.217$ and $\chi=1$. Pink and light blue spheres
denote the attractive and repulsive parts of Janus dumbbells, while magenta spheres 
are the spherical particles.}
\label{fig:1}
\end{figure}

At low temperatures, the system splits into two subsystems. One of them
contains a single, big ``cluster'' of lamellary ordered particles, while the second
involves extremely rarefied, gas-like phase.
In figure~\ref{fig:1}a we show a part of the lamellar structure that was
formed at $T^*=0.2$. Note that for the assumed 
cut-off distances of the interaction potentials,
the structural ordering in the system is expected to appear at quite low
temperatures.

The lamellar morphology of the dense phase (figure~\ref{fig:1}a) results from
the attractive interactions in the system, especially from the attraction between
pairs $2$ and $3$. Bilayers of dumbbells that are
formed due to attractive interactions between centers ``1'' 
are ``glued'' together via spherical particles attracted by repulsive parts of the Janus dumbbells. The spherical particles form one-layer stripes and
the entire structure consists of a staggered Janus bilayer and a single layer of spheres.
The two-dimensional structure factor from figure~\ref{fig:1}b) perfectly captures the lamellar nature of the formed phase.
Note that the two-dimensional structure factor was computed using
the method outlined in our previous work~\cite{j7}.

At $T^*=0.2$, almost all the particles in the system are involved in the formation of the lamellar phase.
Only incidentally some single particles of both kinds
have appeared outside this phase, indicating that the gas phase density
in equilibrium with the lamellar phase is almost zero. The nominal
reduced density of the lamellar phase evaluated from the block analysis~\cite{block}
is $\rho^*_L=0.425\pm 0.08$ and its composition is the same as the assumed starting composition, $\chi_L=1$.

We have checked that the lamellar phase
exits up to $T^*\approx 0.35$, while at higher temperatures, the system becomes
a disordered, mixed fluid. Examples of the snapshots illustrating the
temperature evolution of the system are shown in the Appendix, cf. figure~\ref{SI1}. In this Appendix, we also enclosed
the plot of the two-dimensional structure factor (figure~\ref{SI2}) at $T^*=0.4$ that
corresponds to a fluid-like structure.

\begin{figure}[!ht]
  \centering
\includegraphics[scale=0.35]{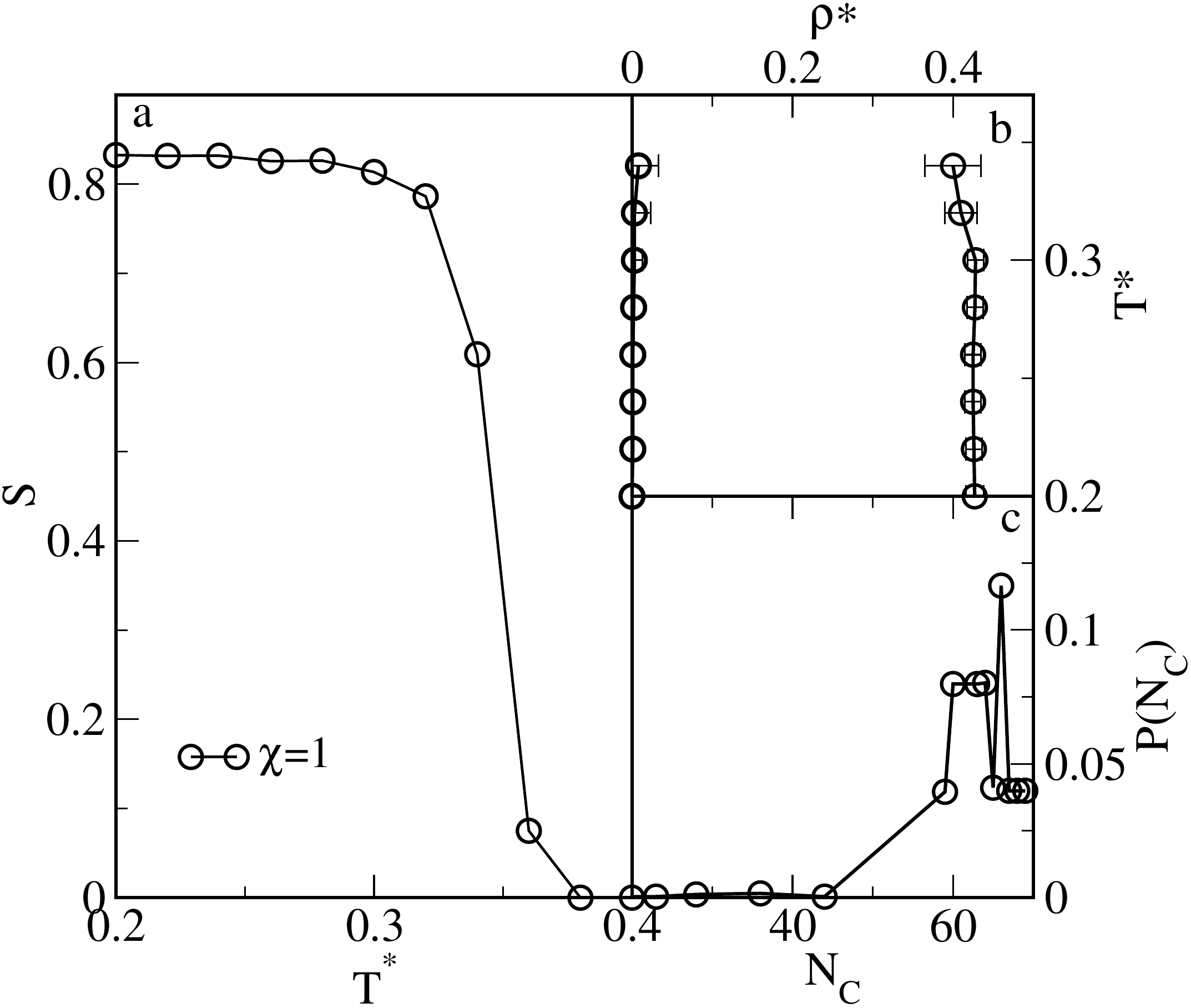}
\caption{ 
Nematic order parameter vs temperature (panel a),
changes of the reduced density of dense and rarefied phases
with temperature (panel b) and the distribution of the cluster
sizes at $T^*=0.3$ (panel c).}
\label{fig:2}
\end{figure}

Figure~\ref{fig:2}a shows the dependence of the order parameter, $S$,
on temperature. Up to $T^*\approx 0.35$, the values of $S$ are almost constant, but
at higher temperatures, they drop to zero indicating that the lamellar
order disappears. Indeed, we observed that at temperatures higher than $\approx 0.35$
the single lamellar cluster splits into several smaller ones.
Figure~\ref{fig:2}b illustrates
the temperature dependence of the density of dense and rarefied phases obtained
from the block analysis.  As in the case of the temperature dependence of $S$,
the data resulting from
the block analysis  indicate  that the separation of the system
into two phases disappears at temperatures higher than $\approx 0.35$.

Figure~\ref{fig:2}c displays
the distribution of the cluster sizes formed by Janus dumbbells at $T^*=0.3$.
The average cluster
size is equal to $\langle N_c \rangle\approx 64$. The tendency to form clusters for the system
with $\chi=1$ vanishes at high temperatures, cf. the Appendix, figure~\ref{SI3}.

Finally, we show examples of the radial distribution functions between the
attractive centers of Janus dumbbells, $g_{11}(r)$, and between spherical species, $g_{33}(r)$.
The results are at two temperatures. At $T^*=0.2$, the lamellar phase exits, while
at $T^*=0.4$ the system contains 
 a mixed disordered fluid. At the low temperature of $T^*=0.2$, both
distributions point to the existence of the order in the system. Interesting is the 
shape of $g_{33}(r)$. The latter function exhibits  ``decaying periodic'' character.
The distance between consecutive periodically recurring patterns corresponds
to the distance between consecutive one-layer stripes of spherical particles.
The inner structure of each pattern results from localization of spherical
particles within each stripe. At $T^*=0.4$, however, the functions
$g_{11}(r)$ and $g_{33}(r)$ are characteristic of a mixed, liquid-like phase.
\begin{figure}[!ht]
  \centering
\includegraphics[scale=0.35]{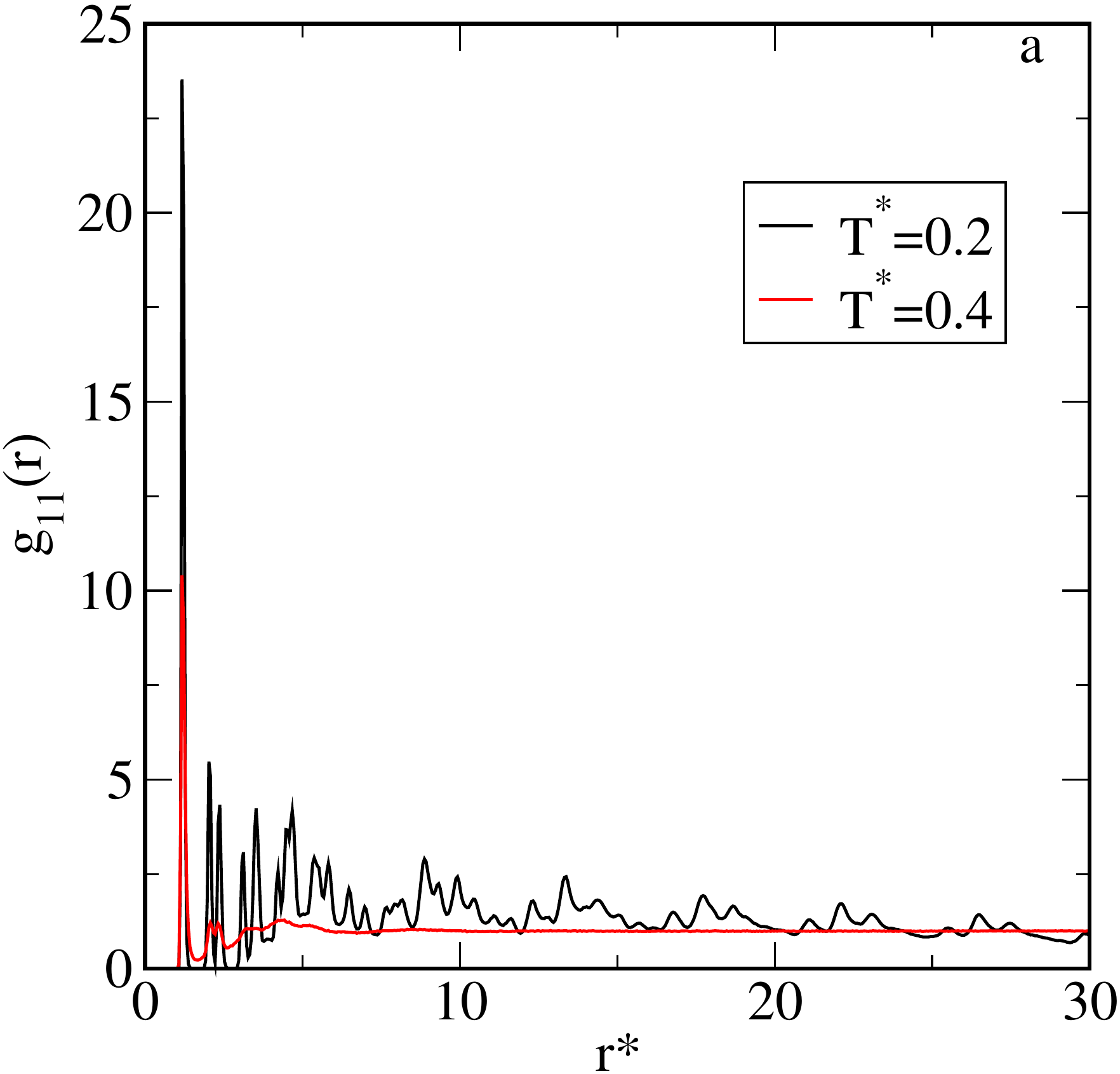}
\includegraphics[scale=0.35]{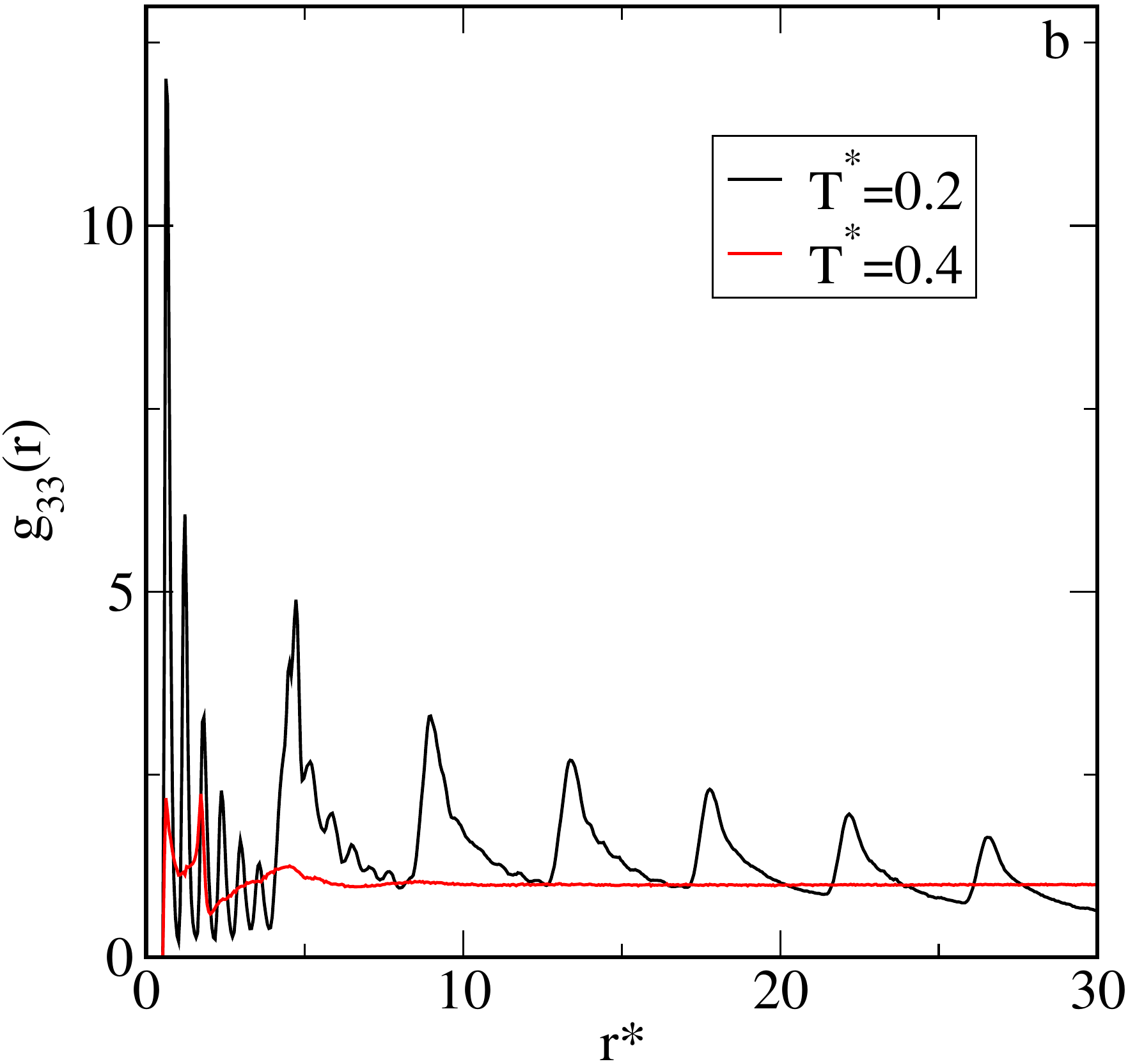}
\caption{ 
(Colour online) Radial distribution functions $g_{11}(r)$ (panel a) and
$g_{33}(r)$ (panel b) at two temperatures, $T^*=0.2$ and 0.4.
The nominal density is $\rho^*=0.217$ and the nominal values of $\chi$ is 1.}
\label{fig:3}
\end{figure}

\begin{figure}[!ht]
  \centering
\includegraphics[scale=0.18]{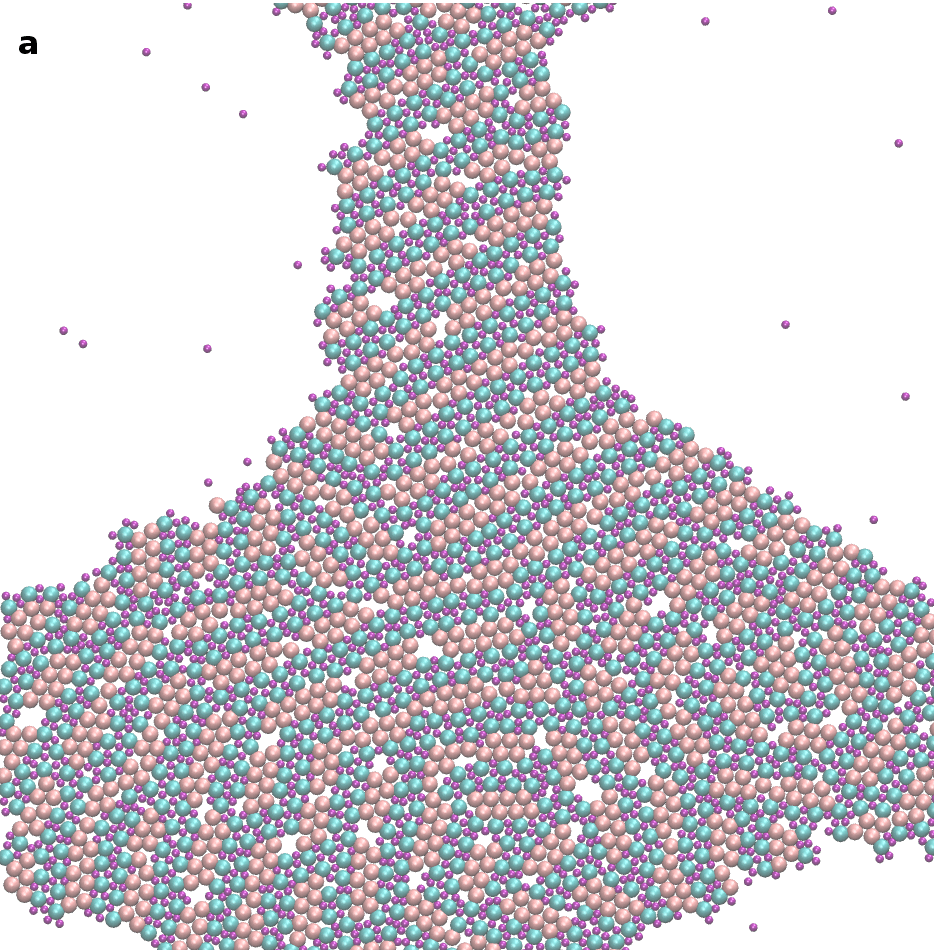}
\includegraphics[scale=0.18]{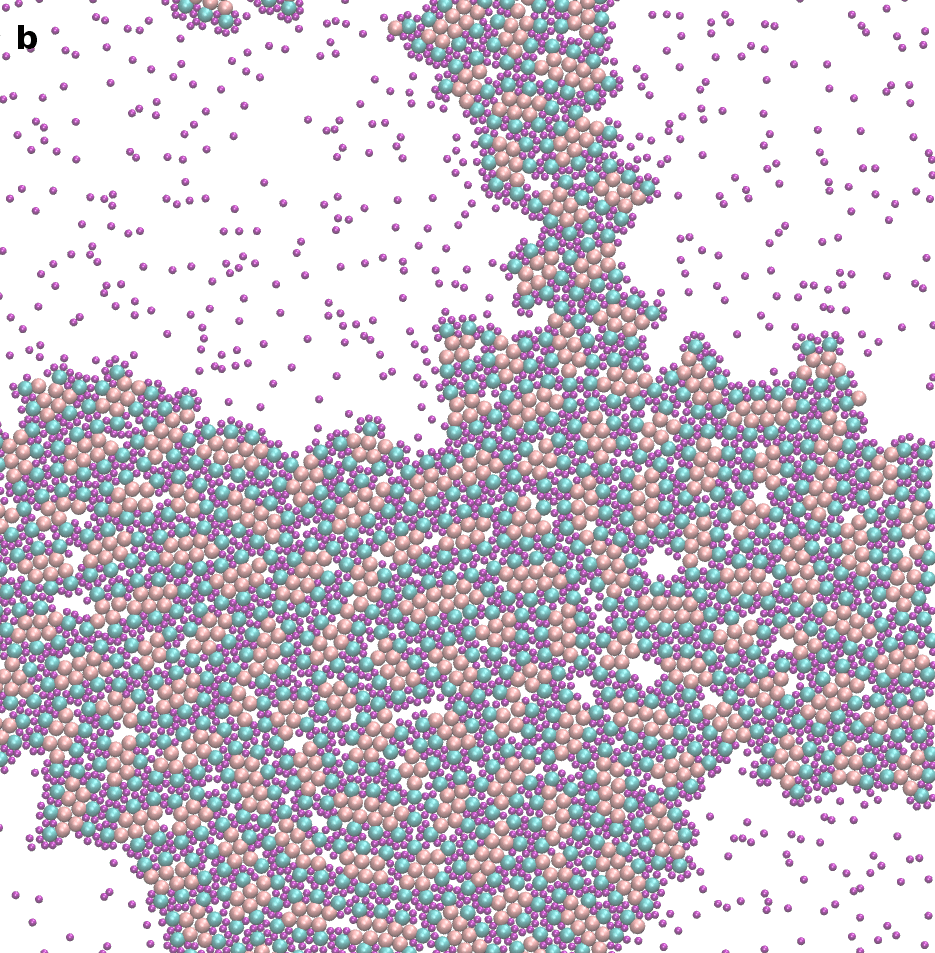}
\caption{ 
(Colour online) Snapshots of configurations in the systems with $\chi=2$,
$\rho^*=0.219$ (panel a) and with $\chi=4$ and $\rho^*=0.214$
(panel b) at $T^*=0.2$.}
\label{fig:4}
\end{figure}

\begin{figure}[!ht]
  \centering
\includegraphics[scale=0.35]{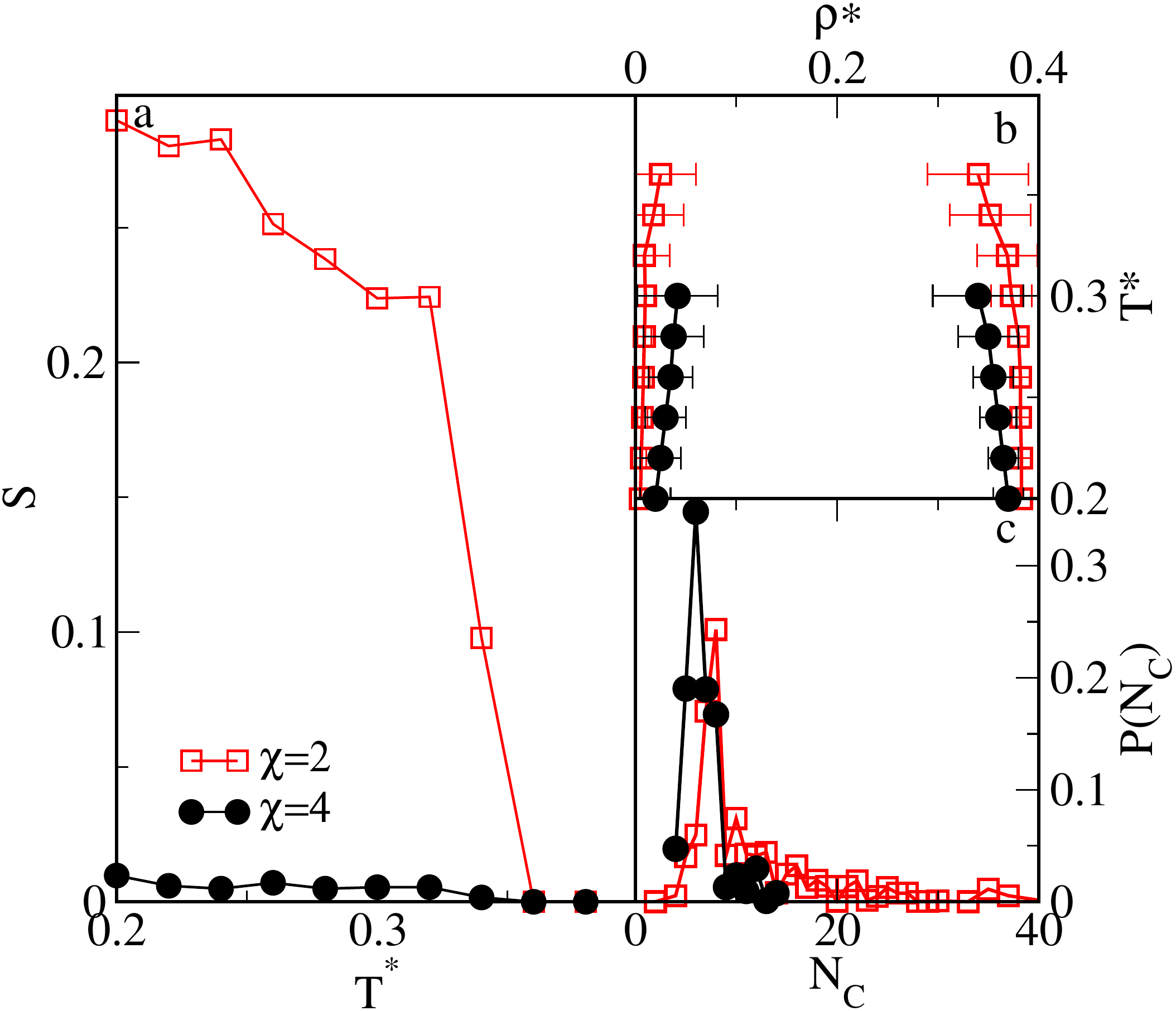}
\caption{
(Colour online) The order parameter  $S$ vs temperature (panel a),
changes of the reduced density of dense and rarefied phases
with temperature (panel b) and the distribution of the cluster
sizes at $T^*=0.4$ (panel c). The nominal densities $\rho^*$ are 0.219
for $\chi=2$ and 0.214 for $\chi=4$.}
\label{fig:5}
\end{figure}

We now proceed to the cases of systems with $\chi=2$ and $\chi=4$. 
The nominal densities
in these systems were $\rho^*=0.219$ and
$\rho^*=0.214$, respectively. In figure~\ref{fig:4} we display the snapshots for both systems
at $T^*=0.2$. 

For $\chi=2$,  the lamellar topography is still preserved at low temperatures
(see figure~\ref{fig:4}a). However, the bilayers of Janus dumbbells and the strips
of spherical particles do not run in parallel, but they bend over longer distances.
This structure can be called ``rippled lamellae''~\cite{i19}.
Although the values of the order parameter $S$ (figure~\ref{fig:5}a) are now lower than 
for $\chi=1$,
the shape of the curves $S$ vs $T^*$ in figures~\ref{fig:5}a and \ref{fig:2}a is
similar and for temperatures lower than $\approx 0.35$ only a single 
``rippled lamellae'' exists
in the system. The density of this phase (figure~\ref{fig:5}b) is lower 
than for the system characterized by
$\chi=1$ (cf. figure~\ref{fig:2}b). Thus, a higher amount of spherical
particles leads to the ``looser'' packing of the dense phase. Consequently,
the distances between Janus particles increase and the cluster size
(cf. figure~\ref{fig:5}c) decreases. The average
cluster size for $\chi=2$ is  $\langle N_c \rangle\approx 11$, thus it is significantly lower than in
the case of $\chi=1$.

In the systems with $\chi=2$,
the total nominal number of spherical particles is two times higher
than the nominal number of the Janus dumbbells.  In the dense phases, the spherical particles try
to surround the repulsive parts of Janus dumbbells. Since spherical particles appear in excess
to dumbbells, the bilayers are twisting. Moreover, some spherical particles
are released from the dense phase. Consequently, the rarefied phase consists almost exclusively
of spherical particles. Figure~\ref{SI4} in the Appendix provides additional insight into disaggregation of the initial large cluster in the system with $\chi=2$. 
Pictures of the two configurations at temperatures of $T^*=0.36$ and 0.38 are shown there.
Despite the slight temperature difference, the configuration
of the molecules at those two temperatures is completely different.

Our calculations have indicated that at $T^*=0.2$, the composition of the dense phase, $\chi_L$,
(evaluated from block analysis) is $\chi_L\approx 1.91$.
With temperature increase, the composition of the dense phase  decreases slightly:
at $T^*=0.3$ it equals $\chi_L\approx1.88$ and at $T^*=0.34$ --- $\chi_L\approx 1.84$.

For $\chi=4$, the dense phase is still formed at low temperatures (figure~\ref{fig:4}b),
but now it
does not exhibit any lamellar order.
The changes of liquid-like phase density with
temperature is shown in figure~\ref{fig:5}b. The block data analysis
has indicated that at temperatures higher than $\approx 0.31$ the
single liquid-like cluster is falling apart into several smaller ones.
The mechanism of disaggregation is the same as outlined previously.

At lower temperatures, the single cluster composition, $\chi_L$, is lower
than the nominal value of $\chi$ and is equal to $\chi_L\approx 0.37$ at $T^*=0.2$,
$\chi_L\approx 0.36$ at $T^*=0.24$ and $\chi_L\approx 0.34$ at $T^*=0.3$.
Similarly to $\chi=2$, the coexisting rarefied fluid is composed
mainly of spherical particles. 

\begin{figure}[!ht]
  \centering
\includegraphics[scale=0.35]{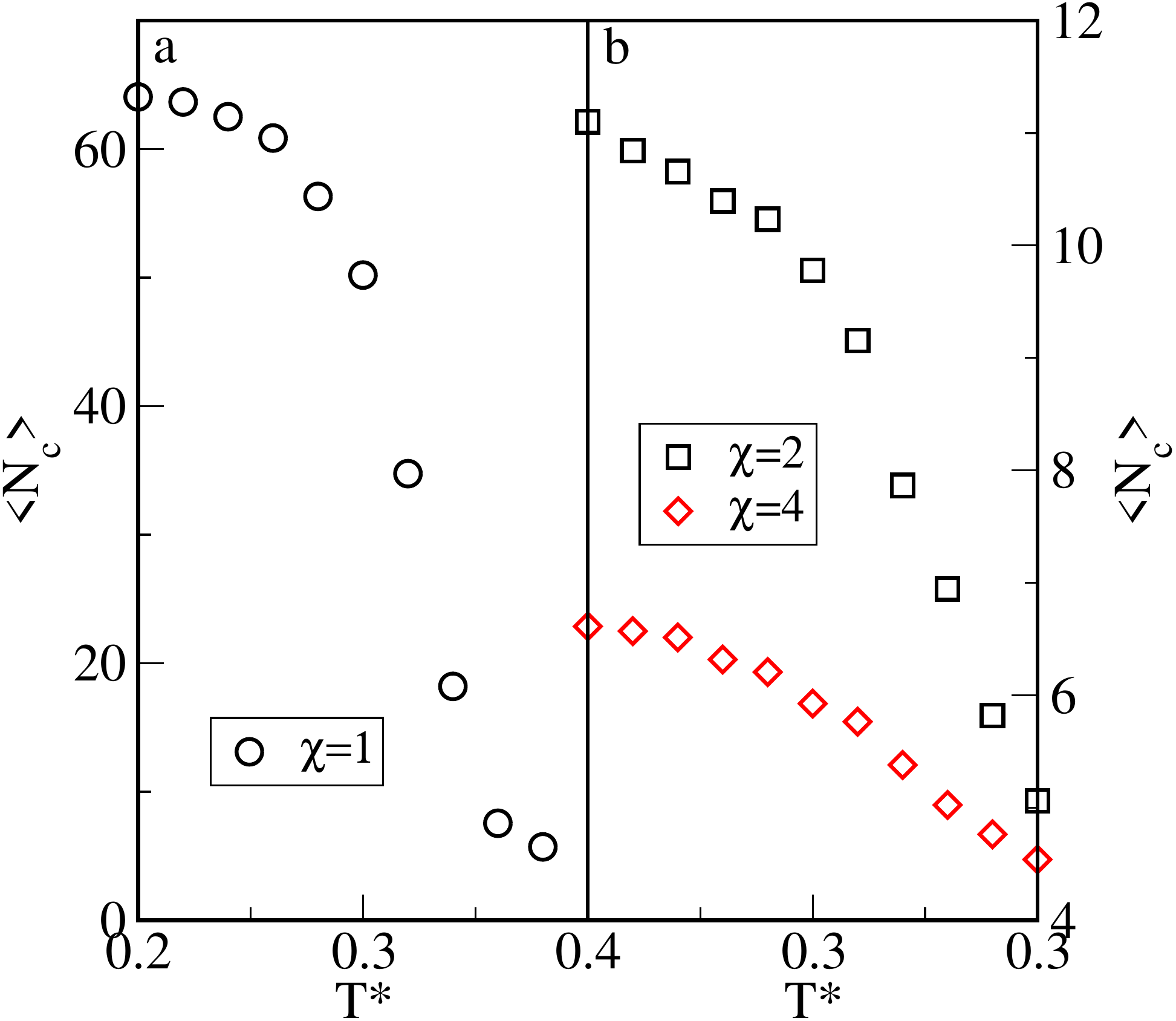}
\caption{ 
(Colour online) The dependence of the average cluster size $\langle N_c \rangle$ on
temperature for the systems with $\chi=1$ (panel a), $\chi=2$ and $\chi=4$
(panel b). The nominal density is 0.217 for $\chi=1$,
0.219 for $\chi=2$ and 0.214 for $\chi=4$. }
\label{fig:6}
\end{figure}

Figure~\ref{fig:6} illustrates the changes of the average cluster
size with temperature. Panel a is for the system with $\chi=1$. In this
case, the drop of $\langle N_c \rangle$ with temperature increase is the most
pronounced and rapid. For the system with $\chi=4$, the increase of temperature causes small changes in the cluster size. Of course, the values  $\langle N_c \rangle$ 
depend on the selected cut-off distance, $R_{JJ,m}$.
We recall that in our calculations this distance was equal to the
distance of the first minimum of the radial distribution function
for the Janus dumbbells.

At high temperatures, the structure of the systems with $\chi=2$ and $\chi=4$  is similar, cf. figure~\ref{SI5} of the Appendix, where we enclosed snapshot of the configurations at
$T^*=0.5$. The radial distribution functions between all spherical species
resemble the plots presented in figure~\ref{fig:3} for the system with $\chi=1$ at $T^*=0.4.$

We now proceed to the system with $d^*=0.5$. The two spheres
forming Janus dumbbells do overlap. Similarly to $d^*=1$,
the calculations have been carried out for $\chi=1$, 2 and 4.

For systems with
$\chi=1$ at low temperatures, only a single cluster composed of domains with non-zero values of the order parameter S appears, cf. figure~\ref{fig:7}a.
Similar situation 
is observed up to $T^*\approx 0.36$, cf. figure~\ref{SI6} in the Appendix. The main difference
between the systems with $d^*=1$ and $d^*=0.5$ is the tendency to
form differently oriented domains in the latter systems, whereas
in the former systems a single cluster of equally oriented stripes  usually appeared. We repeated the
simulation for different starting configurations, and the
tendency to form different domains was observed for all of them.

\begin{figure}[!ht]
  \centering
\includegraphics[scale=0.23]{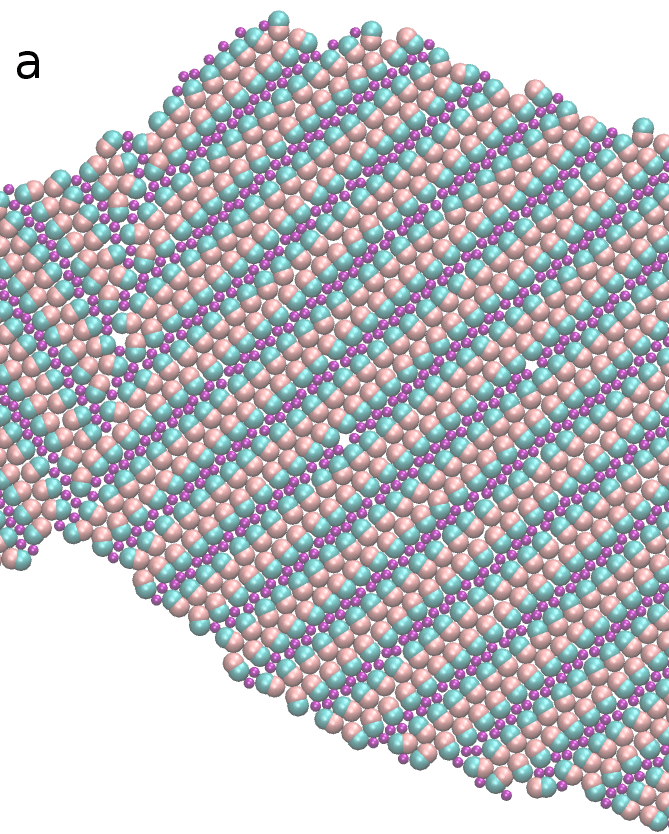}
\includegraphics[scale=0.34]{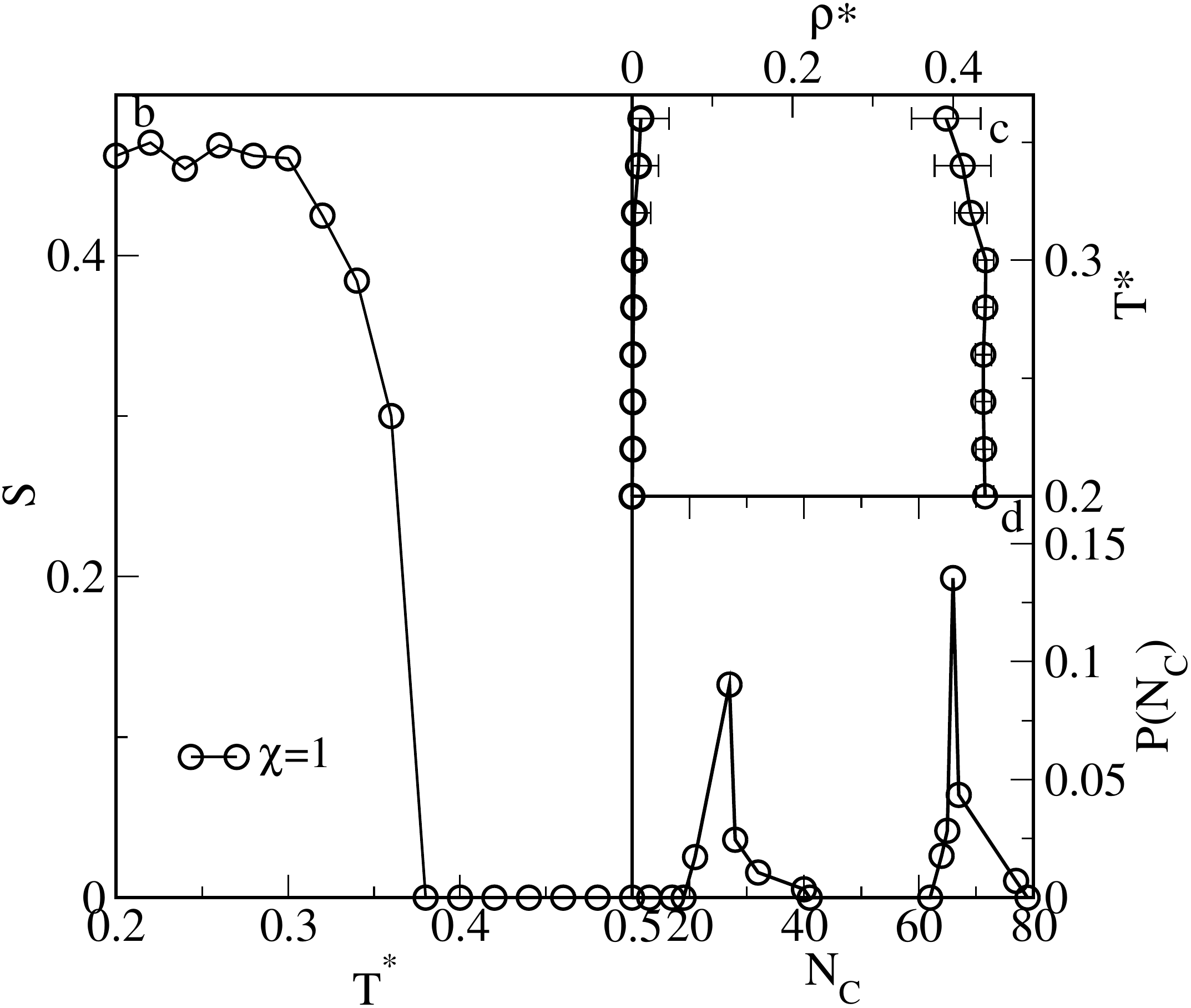}
\caption{ 
(Colour online) Panel a. The snapshot of the configuration for $\chi=1$, $d^*=0.5$ at
$T^*=0.2$. 
The symbols are the same as in figure~\ref{fig:1}a.
Panel b illustrates the changes of the order parameter $S$ with temperature.
Panel c shows the changes of the densities of the coexisting phases
with temperature and panel d displays the cluster size
distribution at $T^*=0.2$. The nominal system density was $\rho^*=0.217$.}
\label{fig:7}
\end{figure}

In figure~\ref{fig:7}b we present the relationship  between the order parameter $S$
and the temperature. The shape of the curve obtained
for $d^*=0.5$ is similar to that for $d^*=1$, cf. figure~\ref{fig:2}a.
However, the value of the order parameter $S$ at a low temperature is almost two times lower
than previously.
This 
is connected with the existence of the domains  mutually oriented at the angle of
$\approx 90^\circ$. 

The changes of the densities of rarefied and dense (lamellar) phases with temperature are
shown in figure~\ref{fig:7}c.  
The density of the condensed phase is now slightly higher than for $d^*=1$.
Similarly to $d^*=1$, the density of the rarefied phase is quite low.
The gas-like phase
consists of nearly equal amounts of spherical and Janus dumbbells.
Thus, for systems characterized by
$d^*=1$ and $d^*=0.5$, the nominal value of $\chi=1$ seems to be optimal for the formation of
ordered, lamellar phases.

Figure~\ref{fig:7}d displays the cluster size distribution $P(N_c)$.
Two peaks of $P(N_c)$ are associated with the existing domains and
for systems with $d^*=0$ and $\chi=1$, we also evaluated the radial distribution functions and
the two-dimensional structure factors. 
Their overall shape and temperature behavior
were quantitatively similar to the system with $d^*=1$.

An increase of the  content of spherical particles to $\chi=2$ in the systems
involving the dimers of elongation $d^*=0.5$
leads to temperature changes similar to $d^*=1$. 
An example of the configuration at a low temperature is shown in the Appendix, cf. figure~\ref{SI7}. 
At low temperatures, the clusters, mainly composed of 6, 7, 8, or 9 Janus
dumbbells, are surrounded by spherical particles. These clusters built larger entities, in which
they are oriented in parallel to each other. Particular larger entities are connected
via spherical particles. 

For $\chi=2$, the number of spherical particles is evidently too large compared to the number of Janus dumbbells
and even at the lowest investigated temperature, $T^*=0.2$, several of them remain outside the big cluster of a dense phase. 
After inspecting several system configurations, we decided not 
to perform a detailed analysis of this case.

\begin{figure}[!ht]
  \begin{center}
\includegraphics[scale=0.23]{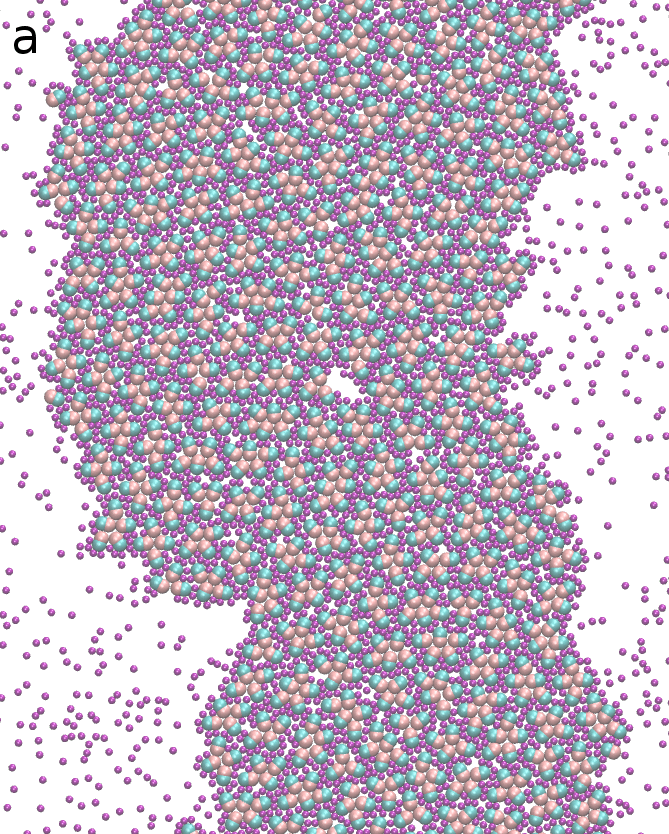}
\includegraphics[scale=0.34]{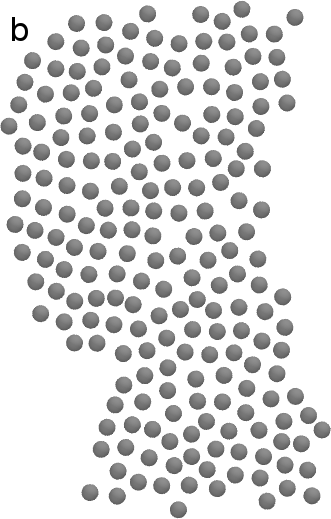}
\caption{ 
(Colour online) Snapshot (panel a) and the configuration of centers of mass of ``elementary clusters'' (panel b)
for the system with $d^*=0.5$, $\chi=4$, at the nominal density  $\rho^*=0.211$ and 
at $T^*=0.2$.}
\end{center}
\label{fig:8}
\end{figure}

\begin{figure}[!ht]
  \begin{center}
\includegraphics[scale=0.21]{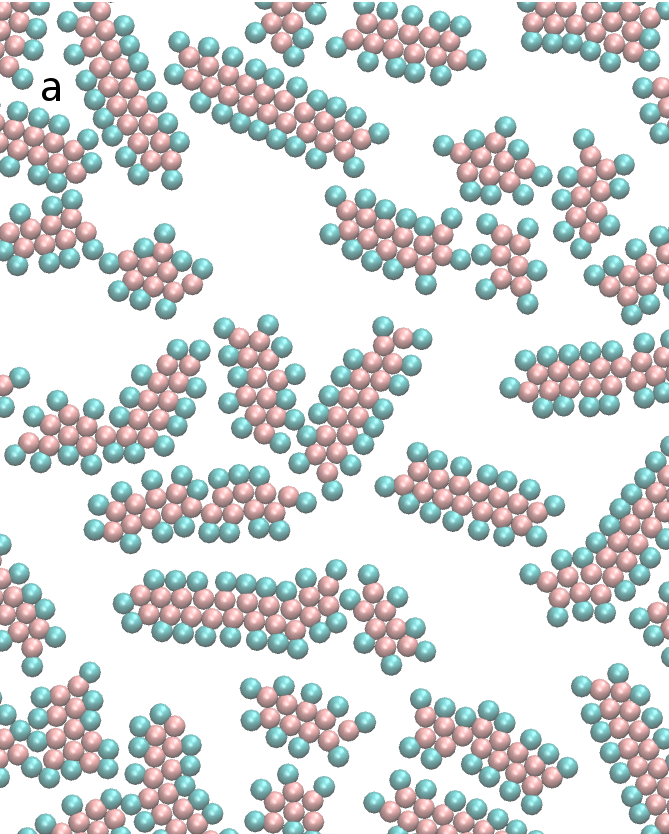}
\includegraphics[scale=0.21]{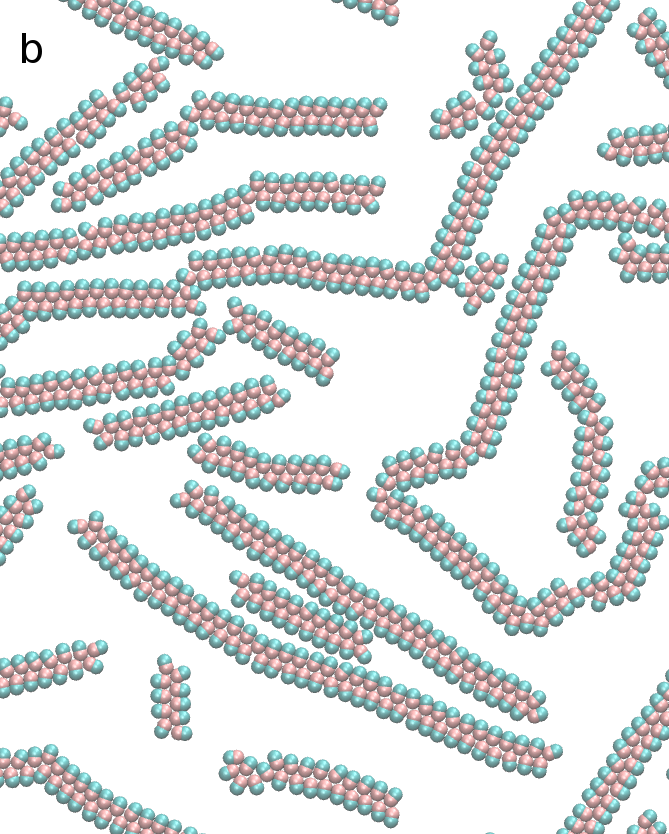}
\end{center} 
\vspace{0.1cm}
\begin{center}
\includegraphics[scale=0.35]{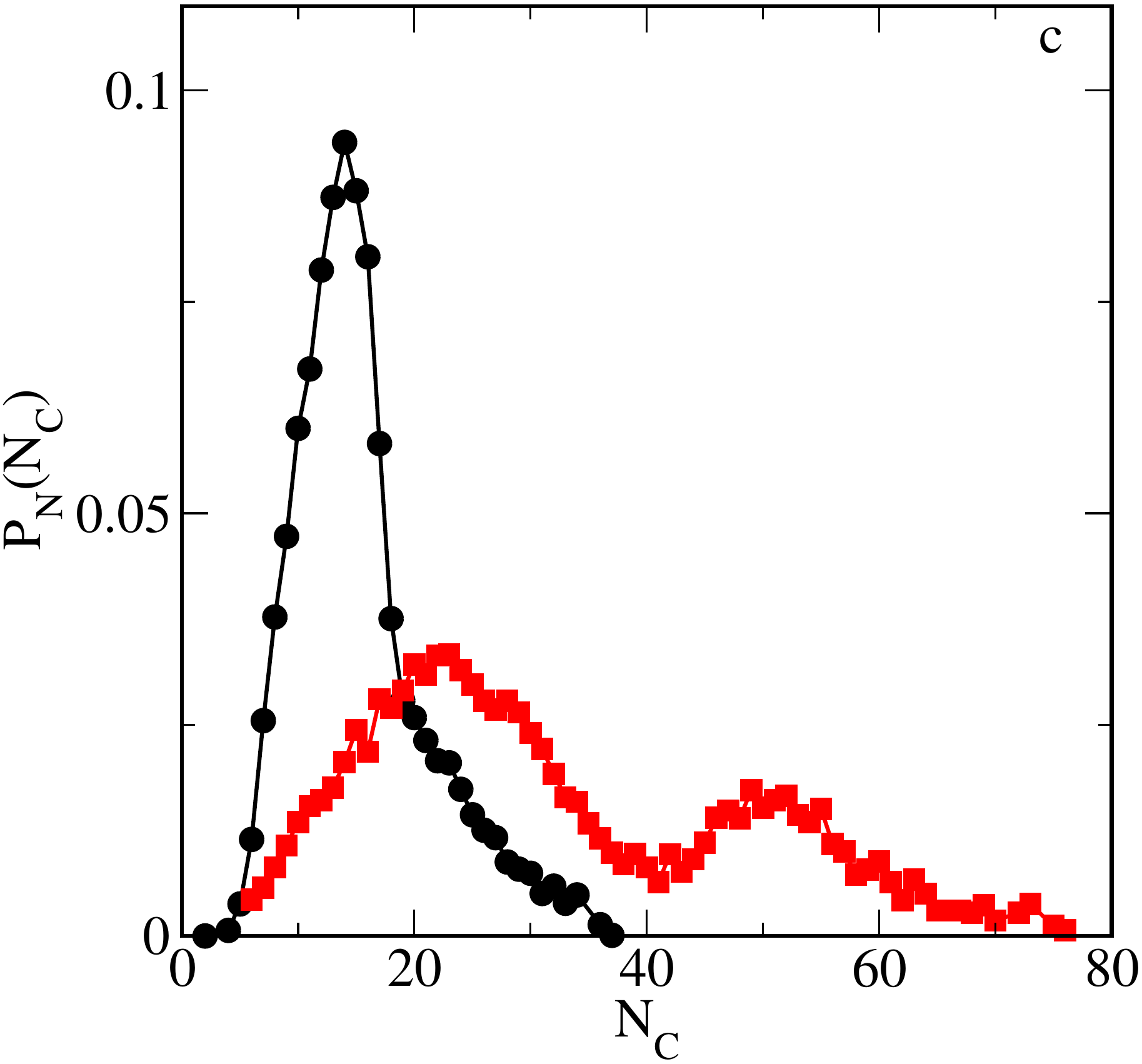}
\end{center}
\caption{ 
(Colour online) Snapshot for the system containing only Janus dumbbells. Panel a is for $d^*=1$, while
panel b for $d^*=0.5$. 
Snapshots show only a part of the entire system. Panel c. The distribution
of the cluster sizes for systems with $d^*=1$ (black line ans circles) and
$d^*=0.5$ (red lines and squares). The nominal density is 0.22 
for both bond lengths $d^*$. The temperature is $T^*=0.2$.}
\label{fig:9}
\end{figure}

With a further increase of $\chi$, the evolution of the structure of the system
follows along the same lines as for $d^*=1$. For $\chi=4$, 
the dense phase is composed of clusters  involving a small number  of Janus particles 
(predominantly 4; we call them ``elementary clusters''),
separated by spherical ones, cf. the snapshot is shown in figure~\ref{fig:8}a.
The size distribution of the clusters is presented in figure~\ref{SI8}a of the Appendix. We see here a single sharp peak at $N_c=4$; there exist some clusters
composed of 3 and 5, and, marginally, of 6 Janus dumbbells.
The density of the condensed phase at $T^*=0.2$ is $\rho^*\approx 0.38$ and its composition
is $\chi_L\approx 0.38$. Both these values are
somewhat higher than those estimated for the system 
with $d^*=1$ (cf. figure~\ref{fig:5}b). 

We have tried to inspect whether the centers of mass (COM) of ``elementary clusters'' are regularly
distributed in the space. This point is illustrated in figure~\ref{fig:8}b, where we displayed
the positions of COMs of particular clusters. By inspecting their configurations we can realize
some traces of hexagonal ordering. However, the analyses of the radial distribution functions
of COM,
(cf. figure~\ref{SI8}b in the Appendix),
as well as two-dimensional structure factors do not discover any pronounced hexagonal
ordering. A dense system behaves like a liquid
of spherical particles with floating
small ``elementary clusters''.

Before proceeding further, we summarize  
the results obtained for mixtures involving Janus dumbbells of the lengths of $d^*=1$ or $d^*=0.5$
and spherical particles. In
both cases, we observed the formation
of lamellar structures. However, the appearance of the ordering of that type is connected
with an appropriate composition of the system. We can say that only for $\chi=1$,
a pronounced lamellar ordering can exist at low temperatures. For any other composition, this
ordering is, at least partially, destroyed by an excessive number of spherical particles, due to their
attraction by the repulsive parts of Janus dumbbells. For shorter bonding distance, $d^*=0.5$,
this effect is more visible. However, one can ask whether the presence of spherical particles is necessary to create this ordering. To answer this question we carried out simulations of the systems containing dumbbell particles alone. 

In figure~\ref{fig:9} we show exemplary configurations of the systems at a
quite low temperature of $T^*=0.2$. For both elongations, i.e., for $d^*=1$ (panel a) and 
$d^*=0.5$ (panel b), we observe the formation of clusters.
The dependence of the distribution of the cluster sizes 
is shown in figure~\ref{fig:9}c.
The size of the clusters
is smaller for $d^*=1$, while for $d^*=0.5$, elongated, spaghetti-like 
clusters appear. Note that the clustering observed in
figure~\ref{fig:9}a resembles the one found in~\cite{i19},
for different interaction potential model.

\begin{figure}[!ht]
\label{fig:AAA}
  \begin{center}
\includegraphics[scale=0.6]{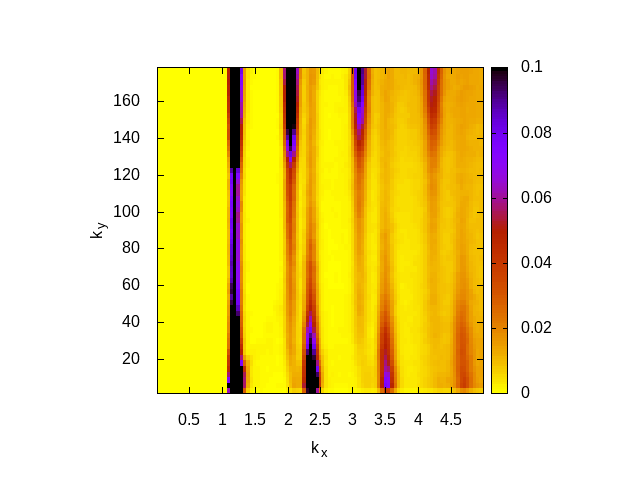}
\caption{\footnotesize \baselineskip=0.5cm 
(Colour online) The two-dimensional structure factor for the system with $d^*=0.5$.
The nominal density is $\rho^*=0.22$ and the temperature is $T^=0.2$.}
\end{center}
\end{figure}

It is also of interest that the clusters
formed by shorter ($d^*=0.5$) Janus dumbbells are partially
ordered in parallel, one relative to the other. Such an effect
is not seen for longer ($d^*=1$) Janus dumbbells. The existence of 
the partial directional cluster
ordering  is 
also reflected by the plot of the two-dimensional structure factor,
cf. figure~\ref{fig:AAA}.

The formation of clusters results from the attraction between
the centers 1 of Janus dumbells. Preserving the geometry of Janus particles and, at the same time,
decreasing the value of the cut-off distance
$r_{11,c}^*$, we could observe the disappearance of the tendency to form clusters. 
The clusters exist since the Janus dumbbells
are stuck together by attractive forces between
their attractive parts. With temperature increase, 
the effect due to the attractive forces becomes less significant 
and the entropic effects lead to the breakup of long clusters into 
smaller parts, cf. figure~\ref{SI9} in the Appendix.

Any attractive center of the Janus dumbbell located inside the cluster
interacts with four attractive centers of other Janus dumbbells.
Ideally, the axes of dumbbells would be oriented perpendicularly
to the long axis of the cluster. Such an ``ideal'' configuration is
easier to be realized by dumbbells with $d^*=0.5$ than 
by more elongated dumbbells, since the repulsive parts of shorter
Janus dumbbells are more effectively ``locked'' by repulsive potentials
$u_{22}(r)$ and $u_{12}(r)$. Consequently, the clusters are longer
for $d^*=0.5$. 

If long clusters are developed in the system, 
they can be treated as new entities, interacting  each with the other
almost exclusively via repulsive
forces.
The behavior of the system of such complex, elongated entities 
can be interpreted in terms of theories of liquid crystals~\cite{frenkel}.
At moderate densities, the parallel orientation of 
linear clusters is more favorable for long clusters. 
However, even at a quite low temperature, the clusters
do not aggregate into bigger structures, due to the lack of
attraction between them. Moreover, for possible development of a lamellar-like
structure of clusters in the system involving only Janus dumbbells,
the nominal density should be much higher than that used in our study.

By comparing the behavior of one-component and the mixed system,
we can conclude that
the lack of spherical particles prevents the formation
of lamellar phases: there are no spherical particles sticking 
clusters together. 
Therefore,
the presence of the spherical particles is
necessary to observe transitions between disordered,
liquid-like, and orientationally ordered, lamellar-type phases.
\newpage

\section{Conclusions}\label{sec:con}

In this work we reported the results of
Molecular Dynamics simulations of the systems
involving Janus dumbbells and spherical particles.
Janus dumbbells were built of two spheres,
connected together by harmonic springs. Two bond length
sizes were investigated. In the first model, the two
spheres were tangentially jointed, whereas in the second model
they partially overlapped. The spherical particles
were two times smaller than the spheres building
dumbbells.
Our goal was to answer the following questions: (i) how the presence of spherical particles influences 
the self-organization 
of Janus particles and (ii) what is the mechanism of the 
formation of self-organized structures. 

The calculations were carried out for a specific model,
according to which the attractive
forces exist between attractive parts of Janus dumbbells
and between repulsive parts of Janus dumbbells and spherical
particles. The forces between all remaining spherical components
were repulsive. Moreover, the range of attractive
interactions was small and,  consequently, the changes
in the structure of the system appeared within
the range of quite low temperatures.

For systems containing Janus dumbbells and spherical
particles, the existence of disordered and lamellary ordered phases
was observed. The formation of lamellar phases was due to 
``gluing'' bilayers composed of Janus particles
by a single layer of spherical particles. Therefore, the presence of
spherical particles is crucial for the development of lamellar
structures. Indeed,
this transition disappeared in the systems involving
no spherical particles. On the other hand,
in the case of mixtures, an
excessive number of spherical particles causes the
destruction of the lamellar structures.

For the highest investigated concentration of spherical
particles, the formation of the orientationally ordered
dense phase is suppressed and the dense phase consists of small clusters
(preferably formed by four, five, or six Janus particles) ``dissolved''
in the fluid of spherical particles. We also checked that this 
phase does not exhibit any pronounced translational order.
This finding is contrary
to that found in~\cite{j7} for mixtures of
spherical Janus-like particles and spherical molecules.

The most pronounced lamellar ordering appeared in the systems containing equal
amounts of Janus dumbbells and spherical particles,
especially when both spheres forming Janus particles partially overlap (the length of 
Janus dumbbells is $d^*=0.5$). This is
undoubtedly connected with the assumed size
of the spherical particles. For a larger
difference of sizes between spheres
forming Janus dumbbells and spherical particles,
the ratio $\chi$ favoring the formation of ordered phases
could be different. 

As we have already stressed, 
an
excessive number of spherical particles causes the
destruction of the lamellar structures. For the highest investigated concentration of spherical
particles, $\chi=4$, we observed  the development of
a dense phase composed of small clusters
(formed by four to six Janus dumbbells) ``dissolved''
in the fluid of spherical particles.
However, the centers of mass of small clusters did not form any
translational ordering. This behavior is contrary
to that found in~\cite{j7} for mixtures of
spherical Janus-like particles and spherical molecules.

One could try to interpret the changes in the studied systems by analogy 
to the behavior of liquid crystals~\cite{frenkel}, 
using the following reasoning:
if the  two dumbbells ``connected'' together
via attractive centers, are supplied with a spherical particle
that can stick to the repulsive center of one dumbbell, then the formed entity can be treated 
as a new ``building block''. This entity   resembles a hard-rod-like particle
and thus the behavior of the system should be similar
to the behavior of hard-rod particles.
However, this interpretation does not seem to
be appropriate since, for liquid crystalline phases, the direction of
long axes of rod-like particles determines the direction of the order.
In our
case, the angle between the direction of axes of dumbbells does not coincide
with the direction of formed layers is non-zero. Moreover, 
the ordering is more pronounced for shorter dumbbells. Therefore, the
explanation of the mechanism of the formation of liquid-crystalline ordered phases
seems to be not appropriate for the description of the changes observed by us.

The model used by us is very simple. However, the  mixture of
asymmetric dimers and disks can also be employed as a model
of amphiphilic molecules (e.g., surfactants) dispersed in an explicit solvent.
In this case, realistic conditions would be those where
the total packing fraction is liquid-like, and disks are more
numerous and much smaller than dimers. In the near future,
we plan to investigate the dynamics of self-assembly in our model,
with specific attention to the nucleation and growth of lamellar
structures.



\section*{Appendix}

\newcounter{appfigure}
\renewcommand{\thefigure}{A.\arabic{appfigure}}
\let\oldcaption\caption
\def\caption{\stepcounter{appfigure}\oldcaption}

This Appendix contains additional figures illustrating the points discussed in the main text.

\begin{figure}[!h]
	\centering
	\includegraphics[scale=0.45]{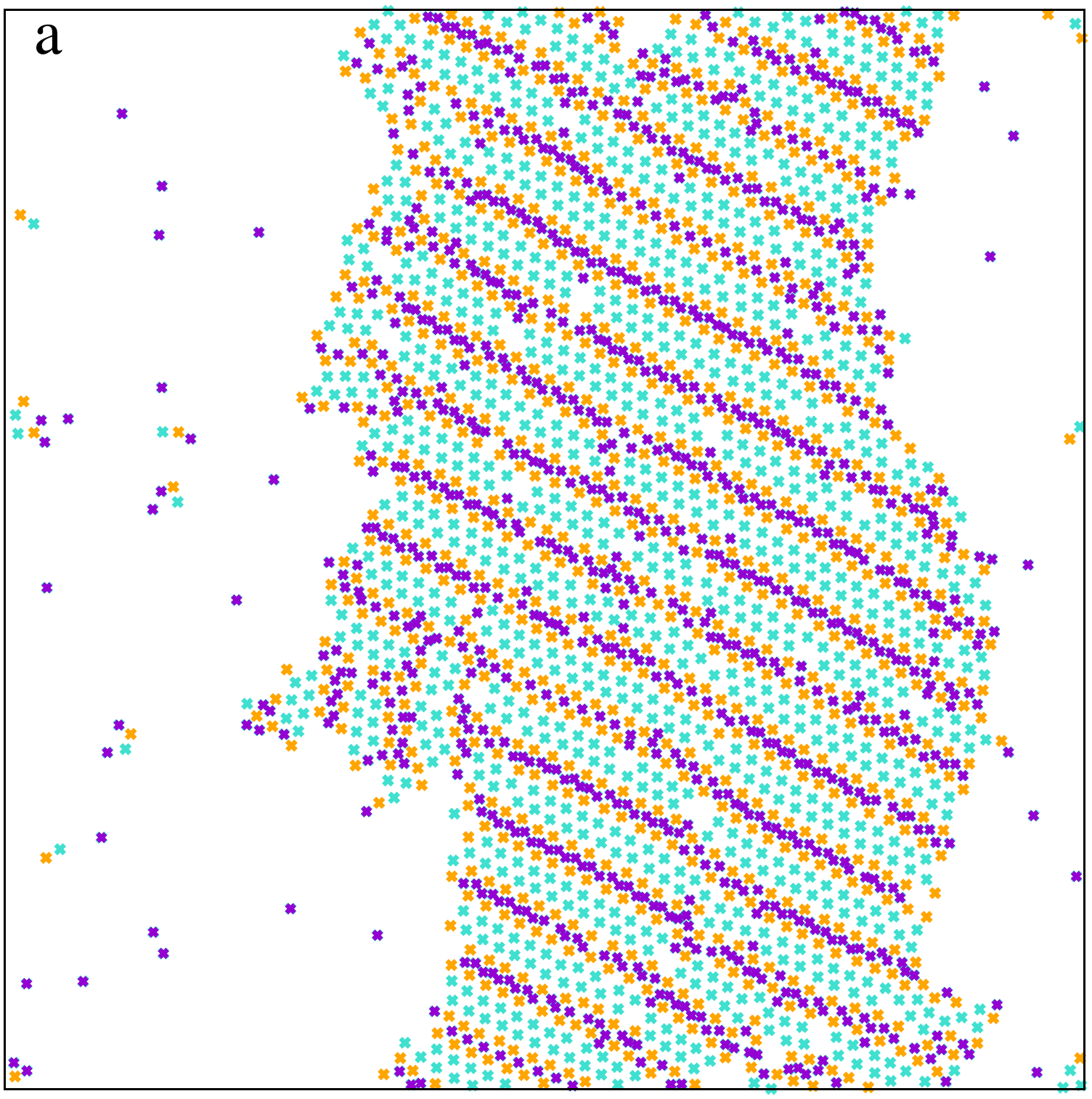}
	\includegraphics[scale=0.45]{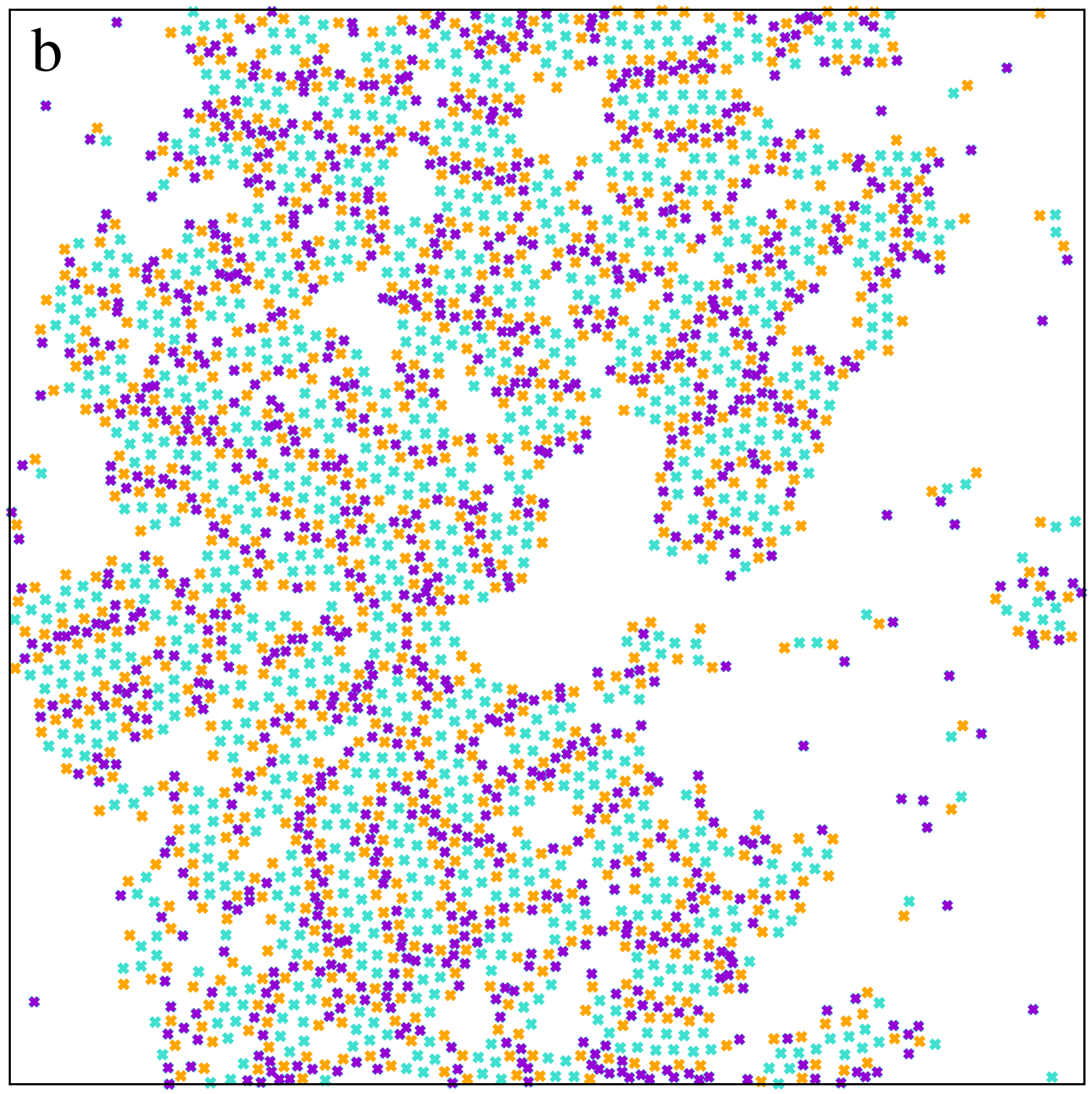}
	\includegraphics[scale=0.45]{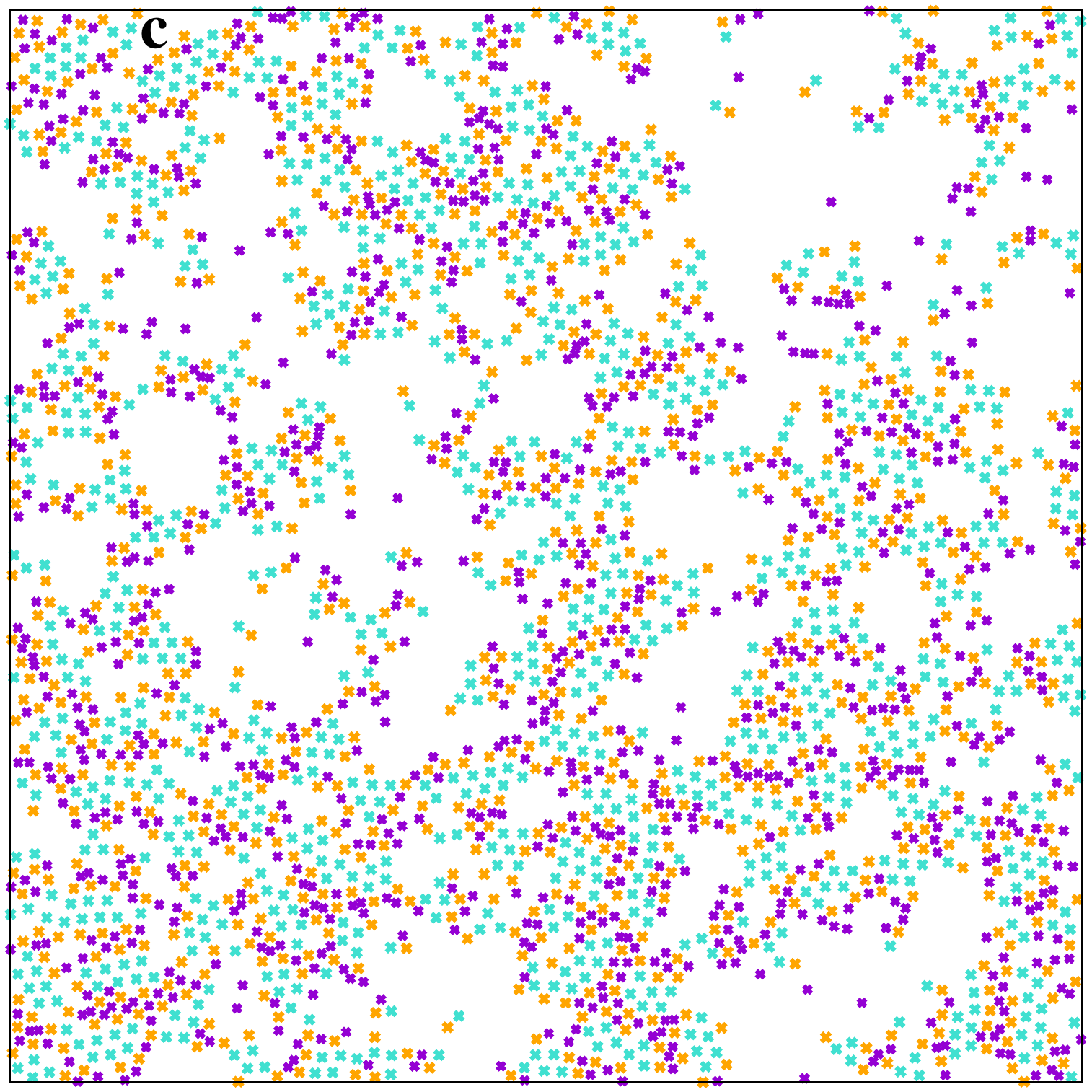}
	\caption{\footnotesize \baselineskip=0.5cm 
		(Colour online) Centers of all species in the system.
		Light blue and orange crosses are for attractive and repulsive 
		Janus particles' components. Violet crosses are the centers of
		spherical particles. The calculations are for $T^*=0.34$ (panel a),
		$T^*=0.36$ (panel b) and $T^*=0.4$ (panel c). The average
		reduced density of the entire system is $\rho^*=0.217$.}
	\label{SI1}
\end{figure}


\begin{figure}[!h]
	\centering
	\includegraphics[scale=0.80]{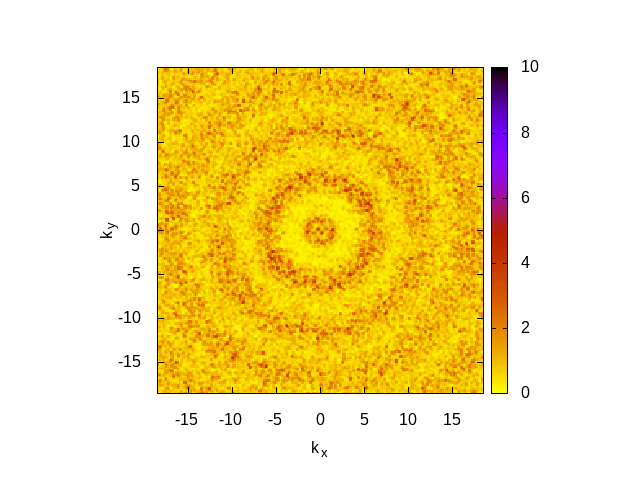}
	\caption{\footnotesize \baselineskip=0.5cm (Colour online) Two-dimensional structure factor for the system at $T^*=0.4$
		and for $\chi=1$ and $\rho^*=0.217$.}
	\label{SI2}
\end{figure}

\begin{figure}[!h]
	\centering
	\includegraphics[scale=0.27]{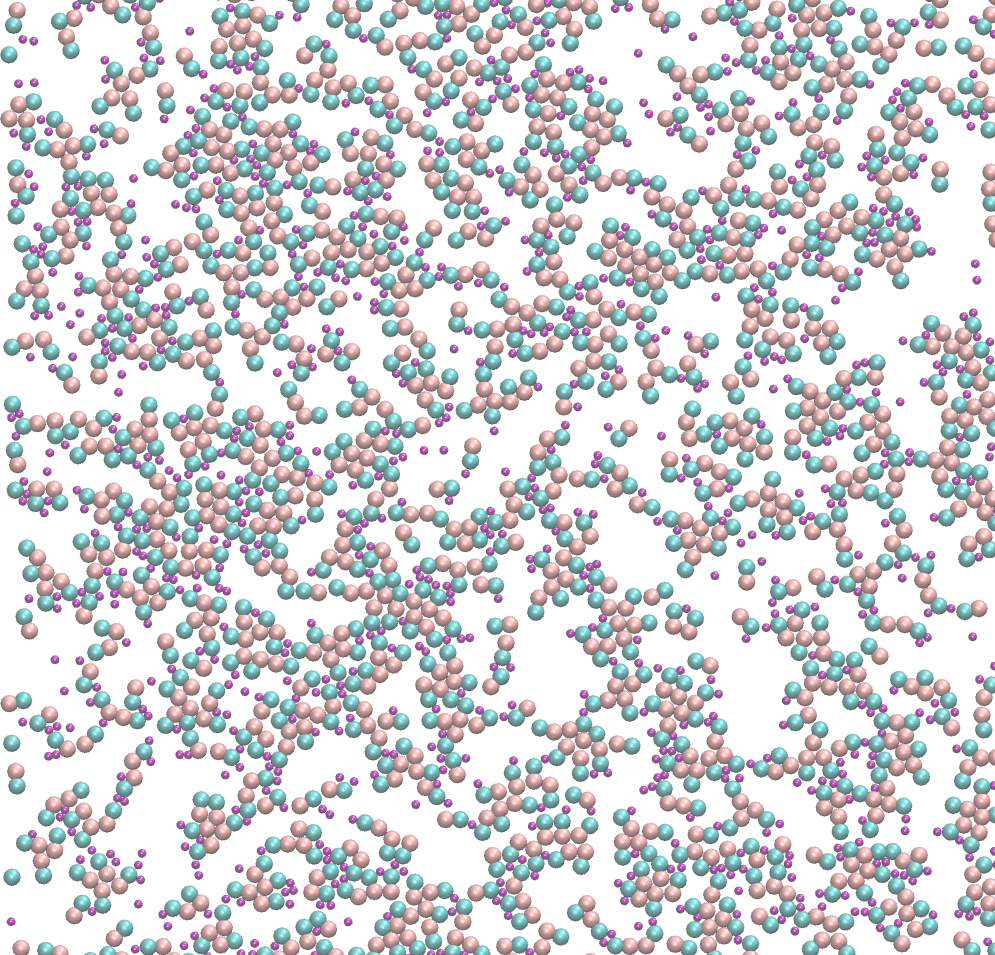}
	\caption{\footnotesize \baselineskip=0.5cm (Colour online) 
		The snapshot of the system with $\chi=1$ at $T^*=0.5$. The symbols
		are the same as in Figure~\ref{fig:1}a.}
	\label{SI3}
\end{figure}

\newpage
\begin{figure}[!h]
	\centering
	\includegraphics[scale=0.45]{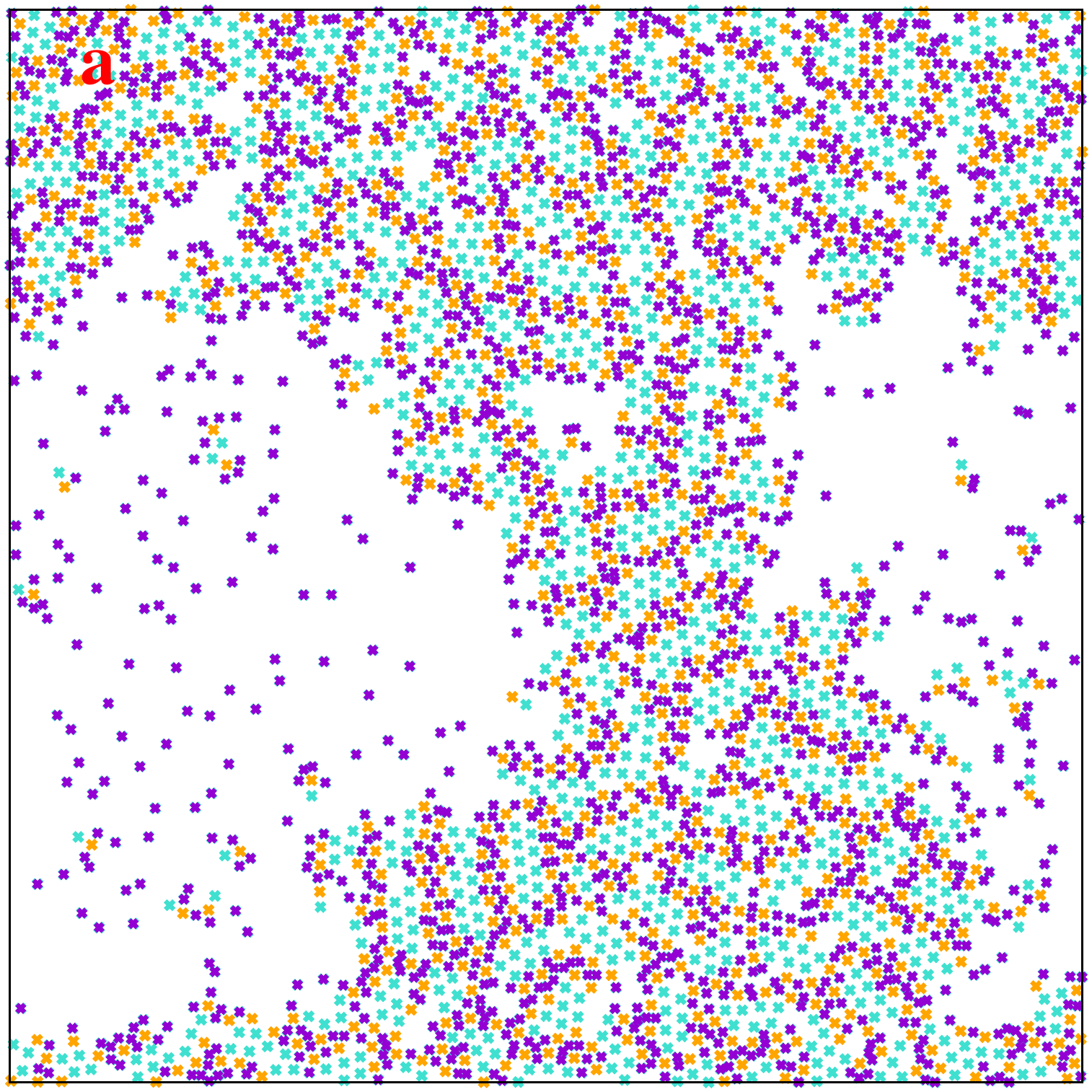}
	\includegraphics[scale=0.45]{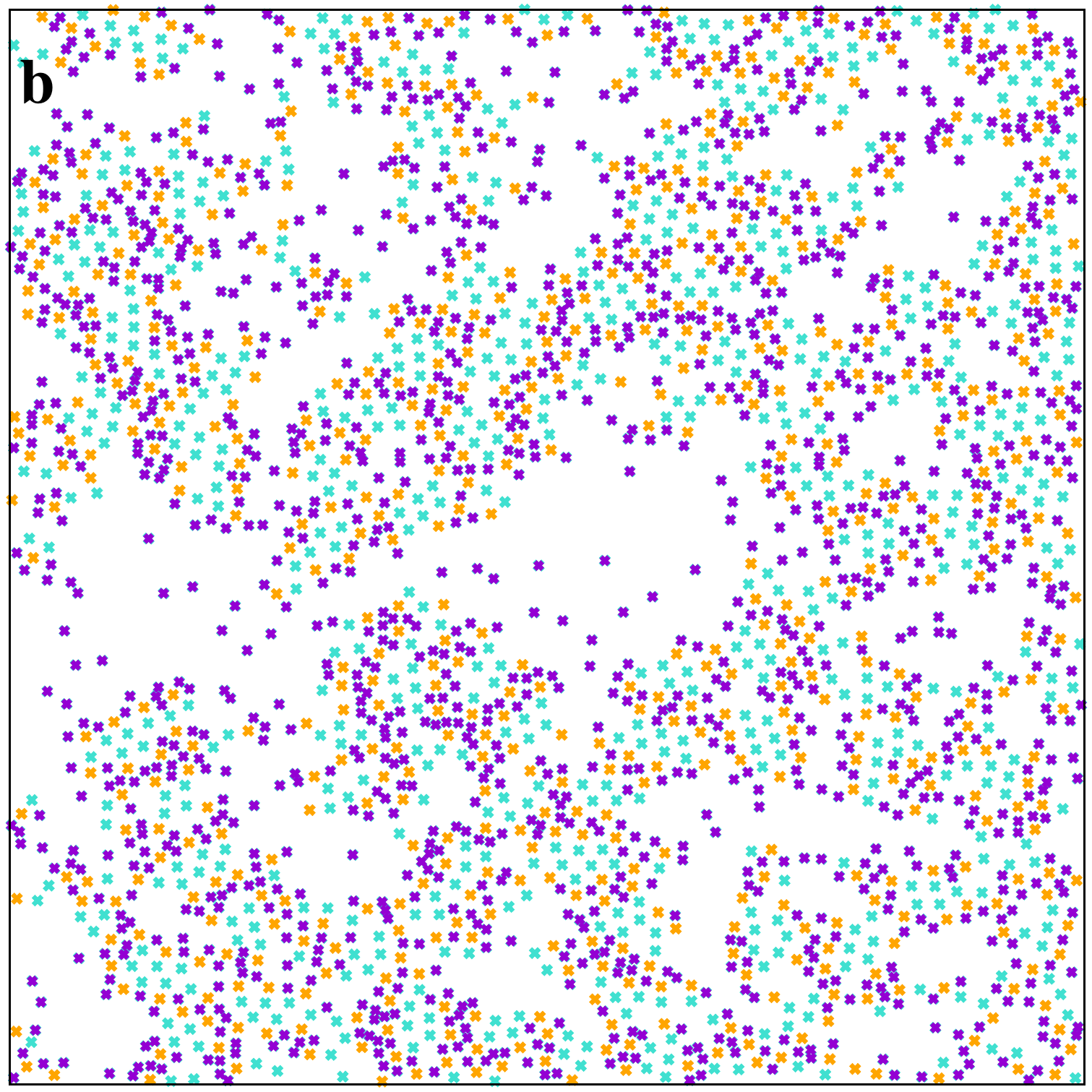}
	\caption{\footnotesize \baselineskip=0.5cm (Colour online) Centers of all species in the system with $\chi=2$.
		Light blue and orange crosses are for attractive and repulsive 
		Janus particles' components. Violet crosses are the centers of
		spherical particles. The calculations are for $T^*=0.36$ (panel a),
		and $T^*=0.38$ (panel b). The average
		reduced density of the entire system is $\rho^*=0.219$.}
	\label{SI4}
\end{figure}

\begin{figure}[!h]
	\centering
	\includegraphics[scale=0.27]{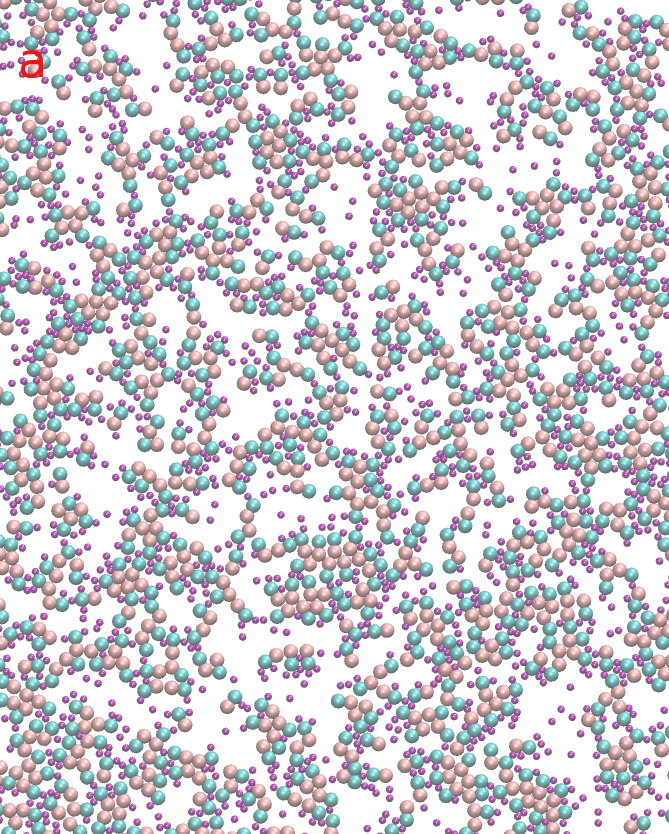}
	\includegraphics[scale=0.27]{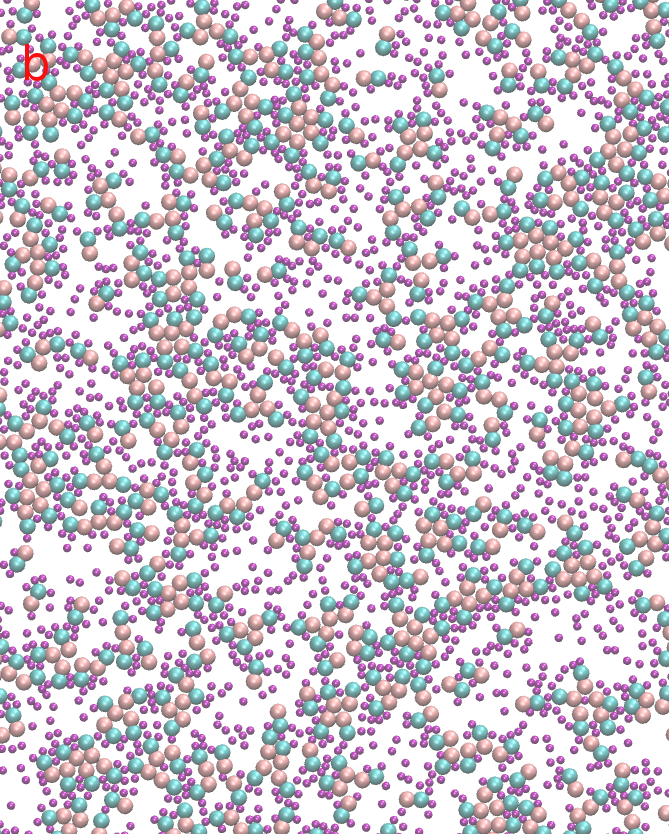}
	\caption{\footnotesize \baselineskip=0.5cm (Colour online) 
		The snapshots of the system with $\chi=2$, $\rho^*=0.219$ (panel a)
		and with $\chi=4$, $\rho^*=0.214$ (panel b).
		Color code is the same as in Figure~\ref{fig:1}a of the main text.
		The calculations are for $T^*=0.5$.}
	\label{SI5}
\end{figure}

\newpage

\begin{figure}[!h]
	\centering
	\includegraphics[scale=0.27]{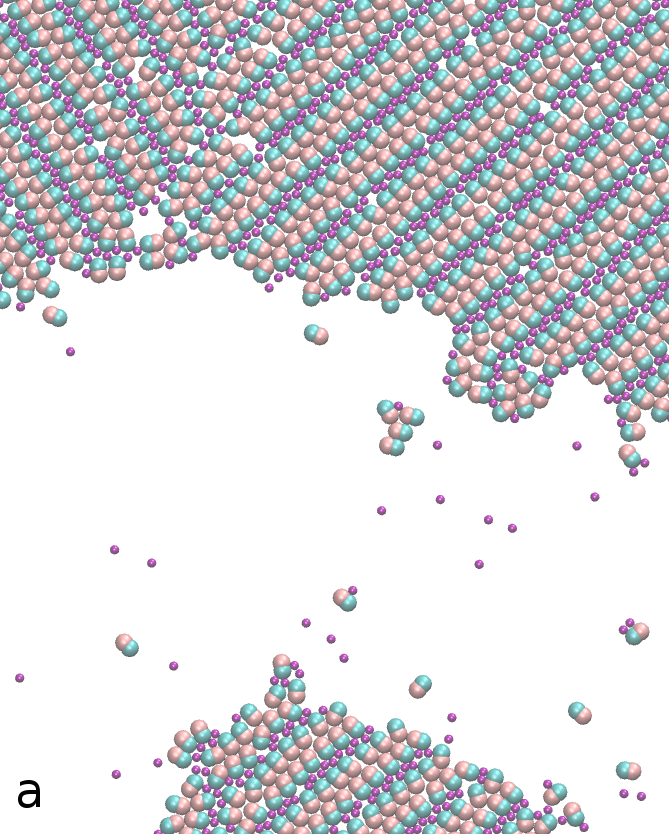}
	\includegraphics[scale=0.27]{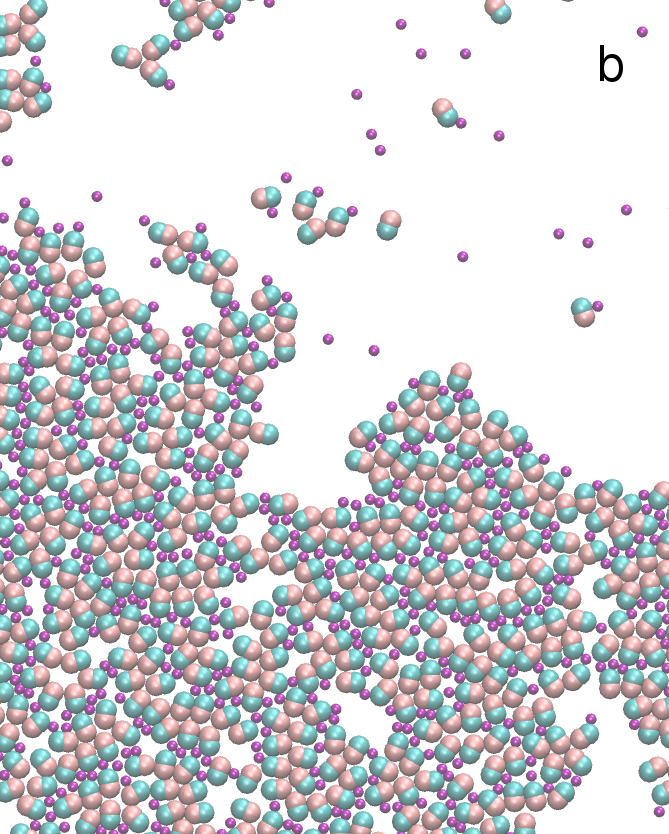}
	\caption{\footnotesize \baselineskip=0.5cm (Colour online) Snapshots for the system with $d^*=0.5$, $\chi=1$ and the starting density of 0.217 at $T^*=0.36$ (panel a) and 0.38 (panel b).}
	\label{SI6}
\end{figure}

\begin{figure}[!h]
	\centering
	\includegraphics[scale=0.27]{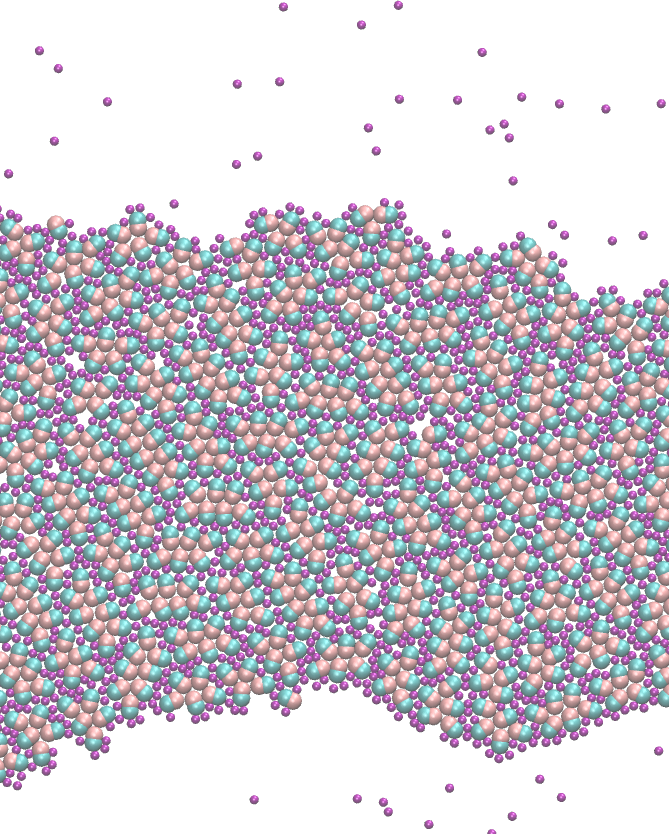}
	\caption{\footnotesize \baselineskip=0.5cm (Colour online) 
		Snapshot for the system with $d^*=0.5$, $\chi=2$ and the starting density of $\rho^*=0.216$
		at $T^*=0.2$.}
	\label{SI7}
\end{figure}
\newpage

\begin{figure}[!h]
	\begin{center}
		\includegraphics[scale=0.37]{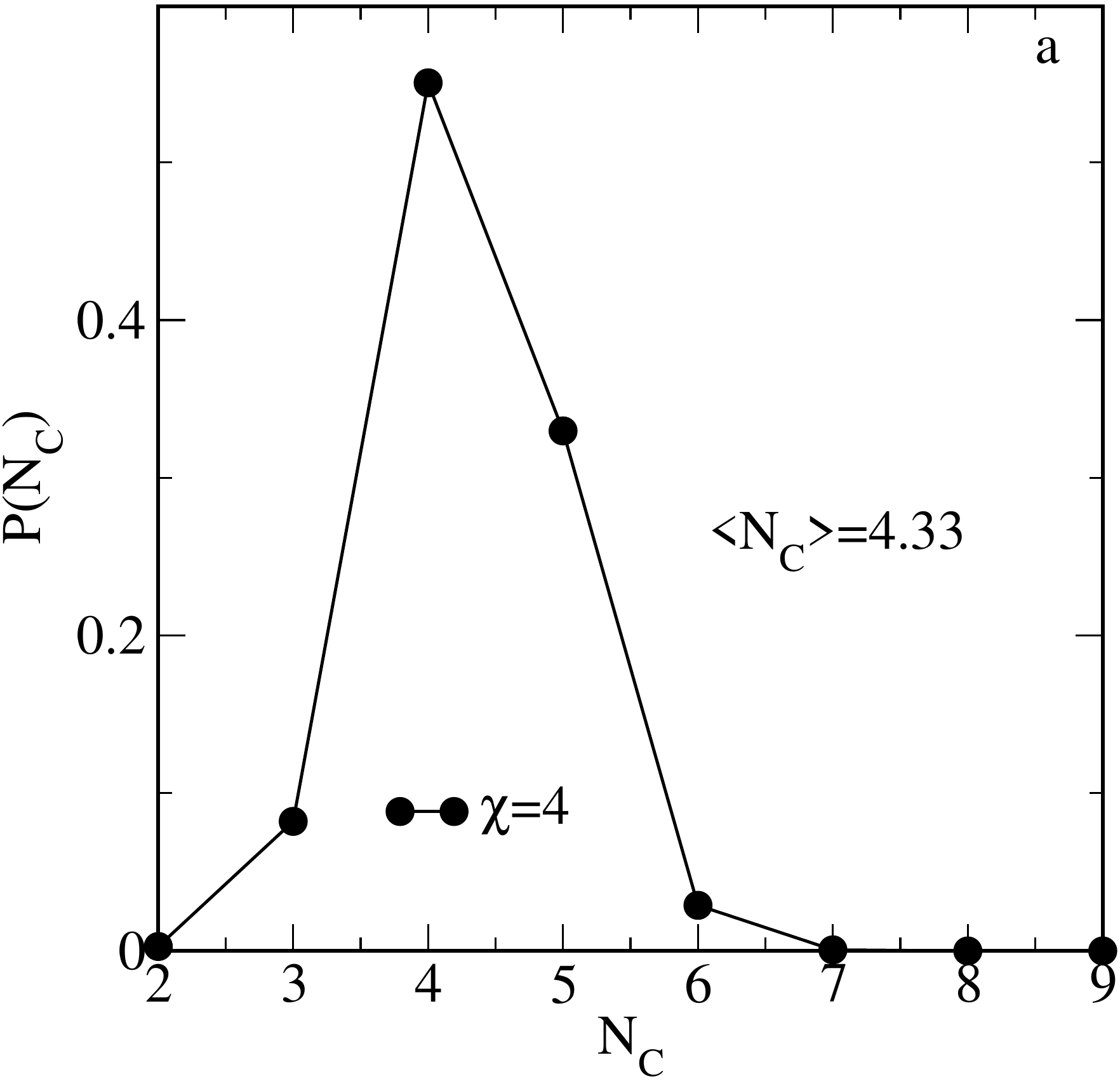}
		\includegraphics[scale=0.37]{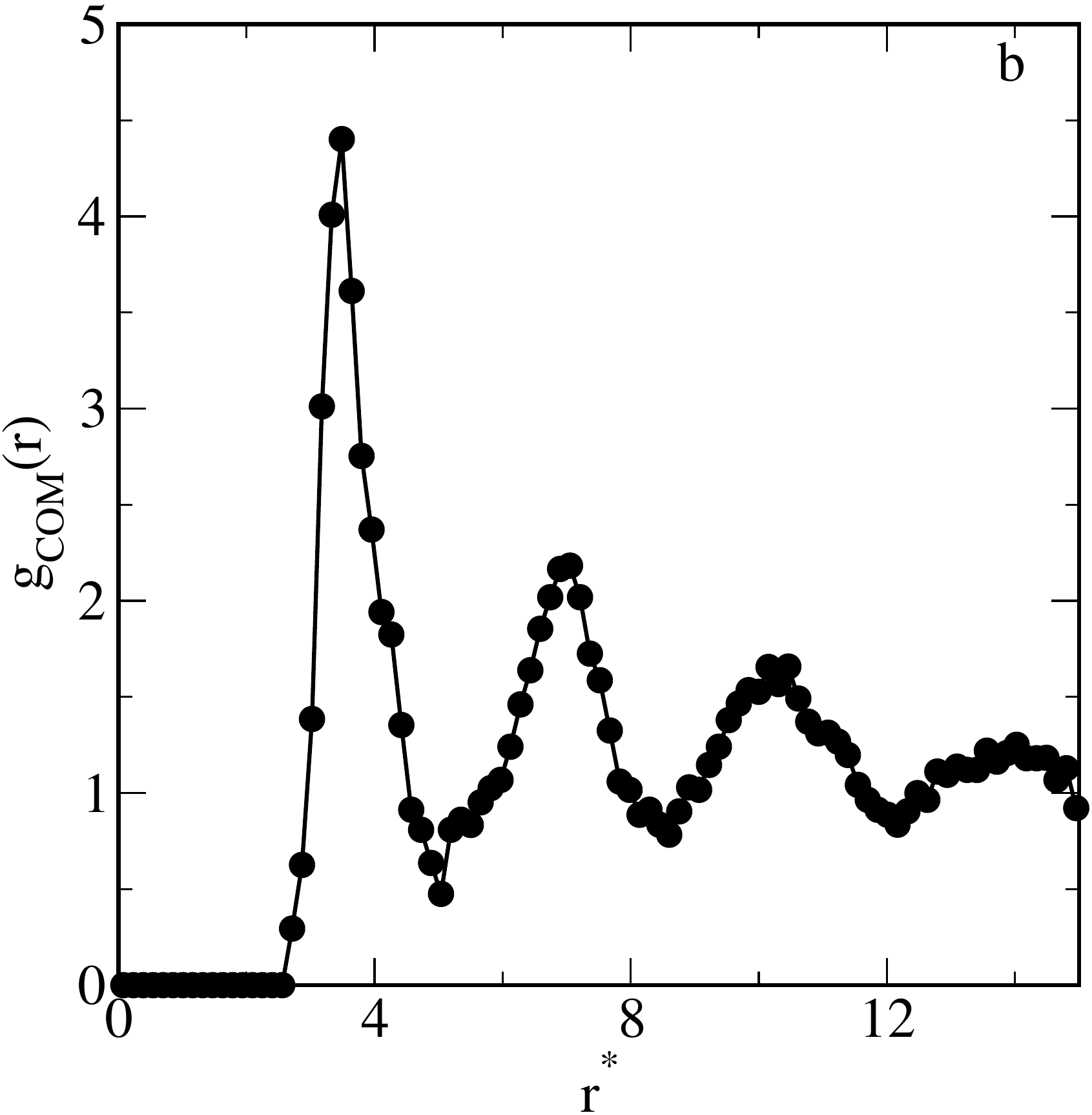}
	\end{center}
\caption{\footnotesize \baselineskip=0.5cm Cluster size distribution (panel a) and the radial distribution
	function of the centers of mass of clusters (panel b) 
	for the system with $d^*=0.5$, $\chi=4$ and $T^*=0.2$.
	The the starting density is $\rho^*=0.211$.}
\label{SI8}
\end{figure}

\begin{figure}[!h]
	\begin{center}
		\includegraphics[scale=0.25]{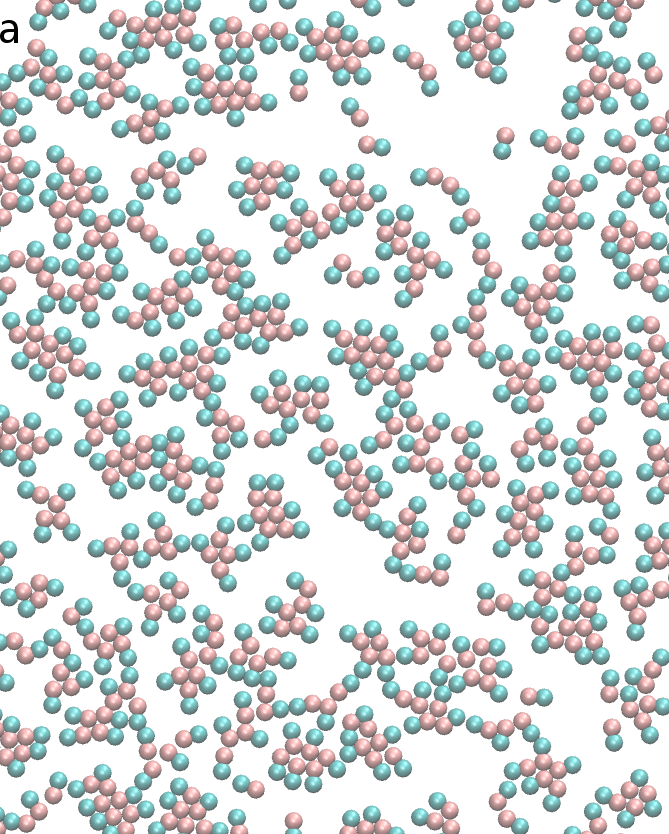}
		\includegraphics[scale=0.25]{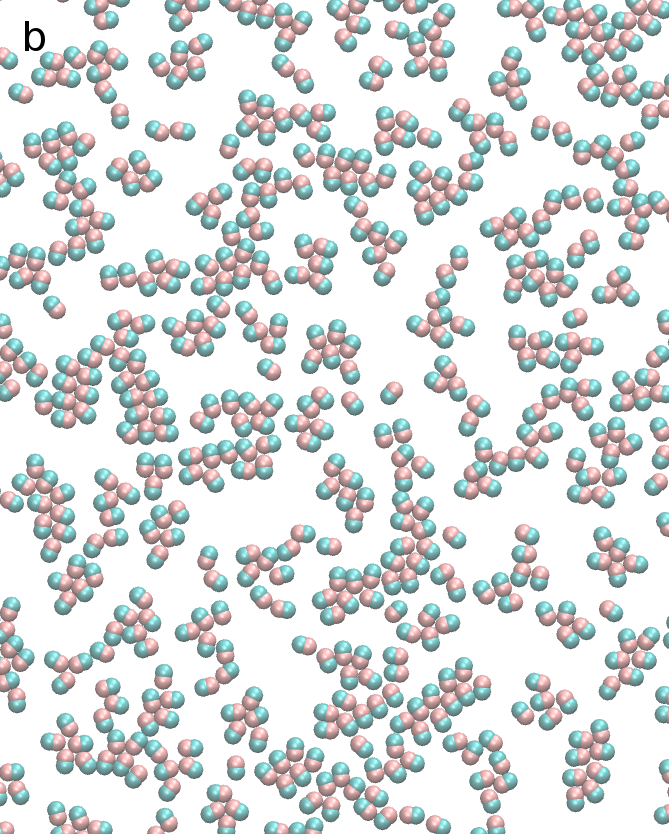}
	\end{center}
\caption{\footnotesize \baselineskip=0.5cm (Colour online) 
	Snapshot for the system containing only Janus particles. Panel a is for $d^*=1$, while
	panel b for $d^*=0.5$. 
	Snapshots show only a part of the system. The starting density is 0.22 and the temperature is $T^*=0.4$.}
\label{SI9}
\end{figure}

\newpage



	\ukrainianpart

\title [Молекулярна динаміка частинок у формі гантелей Януса та сфер]  %
{Молекулярно динамічні дослідження двовимірної системи з частинками у вигляді гантелей Януса та сфер}

\author{Л. Баран, К. Домбровська, В. Жишко, С. Соколовський}

\address{Відділ теоретичної хімії, хімічний факультет університету Марії Склодовської-Кюрі, Люблін 20-031, Польща}

	\makeukrtitle
	
	\begin{abstract}
		
		Проведено детальне молекулярно динамічне моделювання системи двовимірних частинок у формах гантелей Януса та сфер, що перебувають при постійній температурі.
		Гантелі Януса моделювались як дві сфери з мітками 1 та 2, з'єднані гармонічними зв'язками. Сфера 1 вибраної гантелі Януса притягує такі самі сфери інших гантелей, тоді як взаємодія між парами 1-1 та 1-2 є відштовхувальною. З іншого боку, сферичні частинки притягуються центрами 2 та відштовхуються силовими центрами 1 частинок Януса. Показано, що структура орієнтованих  фаз, які можуть виникати в системі, залежить від довжини зв'язків гантелей Януса та відношення між кількістю сферичних частинок та гантелеподібних. Наявність сферичних частинок є необхідною для появи орієнтованих фаз. Для вибраної моделі формування орієнтованих фаз залежить від концентрації сферичних частинок. Однакова кількість частинок Януса та сфер створює оптимальні умови для формування ламеларних фаз.
		
		\keywords моношари, суміші гантелей Януса та сфер, ламеларні фази, молекулярна динаміка, структурні властивості.
	\end{abstract}

\lastpage
\end{document}